\documentclass[12pt,preprint]{aastex}

\usepackage{amsmath,amssymb} 

\shorttitle{Tides in Rotating Planets}
\shortauthors{Goodman and Lackner}

\newcommand{\pd}{\partial}
\newcommand{\ve}{\boldsymbol{\hat e}} 
\newcommand{\vk}{\boldsymbol{k}}
\newcommand{\vkh}{\boldsymbol{\hat k}}
\newcommand{\vn}{\boldsymbol{\hat n}}
\newcommand{\vOmega}{\boldsymbol{\Omega}}
\newcommand{\vv}{\boldsymbol{v}}
\newcommand{\vvg}{\boldsymbol{V}_{\rm gr}}

\newcommand{\vx}{\boldsymbol{x}}
\newcommand{\vxi}{\boldsymbol{\xi}}
\newcommand{\bdot}{\boldsymbol{\cdot}}
\newcommand{\cross}{\boldsymbol{\times}}
\newcommand{\grad}{\boldsymbol{\nabla}}
\newcommand{\diver}{\boldsymbol{\nabla\cdot}}
\newcommand{\cs}{c_{\rm s}} 
\newcommand{\BV}{Brunt-V\"ais\"al\"a\ }

\newcommand{\Rc}{R_{\rm c}}  
\newcommand{\Rp}{R_{\rm p}}  
\newcommand{\Mp}{M_{\rm p}}  
\newcommand{\Mc}{M_{\rm c}}  
\newcommand{\RJ}{R_{\rm J}}  
\newcommand{\MJ}{M_{\rm J}}  

\begin{document}

\title{Dynamical Tides in Rotating Planets and Stars}


\author{J. Goodman and C. Lackner}
\affil{Princeton University Observatory, Princeton, NJ 08544}
\email{jeremy@astro.princeton.edu}

\begin{abstract}
  Tidal dissipation may be important for the internal evolution as
  well as the orbits of short-period massive planets---hot Jupiters.
  We revisit a mechanism proposed by Ogilvie and Lin for tidal forcing
  of inertial waves, which are short-wavelength, low-frequency
  disturbances restored primarily by Coriolis rather than buoyancy
  forces.  This mechanism is of particular interest for hot Jupiters
  because it relies upon a rocky core, and because these bodies are
  otherwise largely convective.  Compared to waves excited at the base
  of the stratified, externally heated atmosphere, waves excited at
  the core are more likely to deposit heat in the convective region
  and thereby affect the planetary radius.  However, Ogilvie and Lin's
  results were numerical, and the manner of the wave excitation was
  not clear.  Using WKB methods, we demonstrate the production of
  short waves by scattering of the equilibrium tide off the core at
  critical latitudes.  The tidal dissipation rate associated with
  these waves scales as the fifth power of the core radius, and the implied
  tidal $Q$ is of order ten million for nominal
  values of the planet's mass, radius, orbital period, and core size.
  We comment upon an alternative proposal by Wu for exciting inertial
  waves in an unstratified fluid body by means of compressibility
  rather than a core.  We also find that even a core of rock is
  unlikely to be rigid.  But Ogilvie and Lin's mechanism should still
  operate if the core is substantially denser than its immediate
  surroundings.

\end{abstract}




\goodbreak
\section{Introduction}
\label{sec:intro}

The discovery of extrasolar planets has sharpened the need for a
predictive theory of tidal circularization and synchronization.  Some
$2\%$ of nearby single FGK stars harbor roughly Jupiter-mass planets
with orbital periods below ten days \citep[and references
therein]{Cumming_etal08}.  The orbital eccentricities of the
shorter-period planets, especially those below four days, are markedly
less than those of the longer period systems, presumably as a result
of tidal dissipation \citep{Rasio_etal96}.  Synchronization is likely
to occur more easily and at longer periods than circularization
because of the smaller moment of inertia associated with the spin as
compared to the orbit \citep{Lubow_etal97}.  For close stellar
binaries, the lack of an adequate tidal theory causes little
uncertainty concerning the internal evolution of the components since
tidal heating contributes negligibly to the luminosity, and since it
is assumed that tides are always sufficient to synchronize the spins
of stars that come into contact with their Roche lobes.  The intrinsic
luminosities of jovian planets due to their gradual contraction and
loss of primordial heat are so small, however, that tidal heating may
be competitive.  Indeed tides have been invoked to account for the
anomalously large radii and low densities, compared to baseline
models, of some planets that are observed to transit their host stars
\citep{Bodenheimer_etal01}.

Most studies of orbital evolution reduce the tidal uncertainties to a
single parameter, the tidal quality factor $Q$, which is inversely
proportional to the tidal dissipation rate, and which one hopes to
calibrate by reference to stellar binaries of known age, such as those
in star clusters, or to the inferred tidal interactions between
Jupiter and its Galilean satellites \citep[and references
therein]{GS66,Mazeh08}.  A difficulty with such pure empiricism is
that $Q$ may depend upon structural details such as composition,
equation of state, rate of rotation relative to the body's density or
to the tidal period, and so on, which differ between the object of
interest and the rather limited set of calibrators.  Indeed there is
some indirect evidence that $Q$ does vary.  As has been pointed out
\citep{Rasio_etal96,Sasselov03,Jackson_etal08}, differences in $Q$
between planets and stars, as well as possible dependence of $Q$ on
the ratio of the orbital to spin frequency \citep{Ogilvie_Lin07}, may
be crucial to the survival of hot Jupiters: since the host stars
rotate subsynchronously \citep{Fabrycky_etal07} and the mass ratio is
large, tidal dissipation within the star would tend to drag the planet
inward.

In order to assess the importance of tidal inputs for hot Jupiters,
furthermore, it is necessary to predict not only the overall rate of
tidal dissipation, but also where within the planet mechanical energy
is converted to heat.  This requires additional assumptions or
knowledge beyond an empirically calibrated $Q$ alone.  For example,
perhaps the best understood and most predictive tidal mechanism, at
least for nonrotating bodies, is the excitation of shortwavelength
g-modes at an interface between convective (isentropic) and radiative
(stratified) regions, originally proposed by Zahn for application to
early-type stars \citep{Zahn70,Zahn75} and first applied to hot
Jupiters by \cite{Lubow_etal97}.  The g-modes would dissipate within
the radiative region since they do not propagate in isentropic
regions. For irradiated planets, as for early-type stars, this means
that dissipation would occur in the outer parts.  On the other hand,
if turbulent viscosity due to convection were effective
\citep{Zahn66}, then the heat would be deposited locally at depth.
Because the thermal timescales and density scale heights differ
greatly between the convective and radiative regions, the g-mode and
turbulent mechanisms would have different consequences for
the planetary radius even if they produced the same $Q$.

As argued by \cite{Goldreich_Nicholson77}, the effective
viscosity of convection is probably very strongly suppressed when the
turnover time of the larger convective eddies exceeds the tidal period
\citep[see also][]{Zahn89,Goodman_Oh97,Penev_etal07}.  Since jovian
planets are very deeply into this regime ($\tau_{\rm conv}\sim
10^3\textrm{yr}$), it is unlikely that convection can be
responsible for $Q$ values as low as are inferred for Jupiter,
$Q_p\approx 10^5-10^6$ \citep{GS66,Peale_Greenberg80}.  While bearing in
mind that phase transitions or other nontrivial microphysics may
contribute to tidal dissipation \citep[e.g.][]{Stevenson80}, we take
the traditional view that the likely alternative is a dynamical tide:
that is, resonant tidal forcing of a low-frequency, short-wavelength
mode or traveling wave at special locations within the planet, such as
the convective-radiative interface already discussed.

Short waves have two basic physical advantages as candidates for a
tidal theory.  First, they are more easily damped than the large-scale
ellipsoidal distortion (``equilibrium tide''), by radiative diffusion
\citep{Zahn75} or by nonlinear breaking
\citep[e.g.][]{Goodman_Dickson98}.  Second, the number of
short-wavelength modes is potentially large, scaling as
$(\Rp/\lambda)^2$ for modes with a fixed azimuthal dependence
$\exp(im\phi)$.  For direct forcing, the frequency of the mode as well
as the azimuthal order $m$ must match those of the tidal potential.
Viewed in the corotating frame of the planet (we assume a uniformly
rotating background state), this frequency is normally much lower than
the fundamental dynamical freqency $\omega_{\rm dyn}\equiv
(3G\Mp/\Rp^3)^{1/2}$, so the modes of interest must be approximately
noncompressive and restored by buoyancy and/or rotation rather than
pressure.  The nomenclature for such low-frequency motions is rich and
bewildering: internal waves, g modes, r modes, toroidal modes, hybrid
modes, Hough modes.  These names distinguish the relative importance
of buoyancy versus rotation and other special properties. In this
paper, we concentrate on modes restored by rotation rather than
buoyancy, which we refer to collectively as ``inertial'' modes or
waves.

All of these nearly incompressible motions have frequencies in limited
ranges controlled by the Coriolis and \BV parameters, regardless of
wavelength.  Local dispersion relations for the wave frequency depend
upon the ratios of components of the three-dimensional wavevector, but
hardly depend upon the wavelength itself, except via viscous or
radiative damping terms.  This behavior is completely different from
that of modes depending upon compressibility (acoustic or p modes) or
selfgravity.  In the ideal-fluid limit, the spectrum of modes that can
be resonant with a tide in the allowed frequency range is therefore
dense \citep{Papaloizou_Pringle81}. But, the tidal potential must have
a finite projection onto a resonant mode in order to excite it. So,
the main challenge for a theory of dynamical tides is to estimate
these projections, also called overlap integrals.

Two recent and independent studies have proposed that inertial modes
may be tidally forced even in the absence of a stably stratified
surface layer, and at rates that approach what is needed to explain
the observationally inferred $Q_p$, provided that the tidal frequency
in the corotating frame is less than twice the rotation frequency;
this is the relevant regime for circularization of a planet that is
already synchronized.  By direct numerical methods,
\citet[hereafter OL04]{Ogilvie_Lin04} calculated the linear excitation
of inertial waves/modes in a compressible spherical annulus surrounding
a solid core.  The current belief is that Jupiter
itself contains $15-30 M_\oplus$ of heavy elements (atomic number
$Z>2$), or $5-10\%$ of the planet's total mass, and also that many or
most of the hot Jupiters may be even more enriched; whether these
``metals'' are concentrated in a distinct core is
uncertain \citep{Guillot05,Guillot_etal06,Burrows_etal07}.  OL04
calculate the tidal dissipation in steady state by balancing the
excitation against an artificial viscous term, but they conclude that
the tidal dissipation due to the excitation of inertial modes, though
varying erratically with frequency in the inertial range (i.e.,
$|\omega_{\rm tide}|<|2\Omega|$), tends towards finite values
corresponding to $Q\sim 10^6$ in the inviscid limit.  This they
explain by the presence in their low-viscosity models of very
short-wavelength disturbances concentrated on a ``web of rays'' in the
poloidal plane.  They find that the intensity of this short-wavelength
response correlates with the size of the core, and they suggest
that it has something to do with closed ray paths (``wave
attractors'') reflected alternately by the inner and outer boundaries
of the spherical annulus.  OL04's interpretation of their own numerical
results seems to have been informed by previous work by 
\cite{Rieutord_Valdettaro97} and by
\cite{Rieutord_etal01}, who studied the linear modes of an incompressible,
slightly viscous liquid in a spherical shell, by a combination
of numerical and analytic methods.  The latter authors demonstrated
the existence of very fine features in the velocity, and they explored the
relationship of these features to critical latitudes and wave attractors,
but they did not consider tidal excitation.
Subsequently, \citet{Ogilvie05} used a simplified
analytical model to argue that wave attractors can absorb energy at nonzero rates
that vary continuously with forcing frequency in the inviscid limit.

The models of OL04 include a stably stratified atmosphere, with an
apparently independent set of ``Hough modes'' excited at the interface
between this zone and the convective region, as previously
demonstrated for rotating massive stars by Savonije and Papaloizou
\citep{Savonije_Papaloizou97, Papaloizou_Savonije97}.  The Hough modes
appear to be much less dependent on the core, and the dissipation
associated with them varies much more smoothly with tidal frequency;
also, these modes extend outside the inertial range, i.e. to
$|\omega_{\rm tide}|>|2\Omega|$.  

Shortly after the work of OL04, Y.~Wu claimed to demonstrate the tidal
excitation of inertial modes in entirely unstratified and coreless
bodies.  In \cite{Wu05a}, exploiting methods developed by \cite{Bryan}
for incompressible rotating bodies, she analyzes the properties of
free (unforced) modes of oscillation in compressible models with
special radial density profiles.  In \cite{Wu05b}, she calculates
spatial overlap integrals between these modes and a quadrupolar
perturbing potential to find $Q\sim 10^9$; she argues that $Q$ may
fall to $\lesssim 10^7$ in more realistic models with a radial density
jump due, for example, to a first-order phase transition.  Using a
different mathematical formalism, but similar underlying low-frequency
approximation, \cite{Ivanov_Papaloizou07} have calculated tidal
excitation of inertial modes in planets on highly eccentric orbits,
with realistic but isentropic equations of state.  This is an
extension of earlier calculations for rotating $n=3/2$ polytropes by
\cite{Papaloizou_Ivanov05}.  Their models are coreless, like those of
Wu, but unlike hers, their tidal excitation is dominated by two
large-scale inertial modes, perhaps because they treat the perturbing
potential as uncorrelated from one periastron encounter to the next,
so that the excitation is non-resonant.

It is clear from the above that considerable progress has been made in
recent years on the dynamical tides of rotating planets, but the
importance of short-wavelength inertial waves remains obscure: in
particular, the circumstances under which they can be tidally excited.
OL04's results indicate that a solid core is
somehow important, while Wu's work suggests that it may not be.  It is
desirable to achieve a semianalytic understanding of the excitation
before making detailed calculations for realistic planetary interiors.
For one thing, since the structure of hot Jupiters is still much less
well understood than that of stars, it will be useful to know what
features of the structure are most important for the tidal problem.
For another, the apparently chaotic variation of tidal torque with
tidal frequency seen in the results of OL04, and to some extent in the
earlier ones of \cite{Savonije_Papaloizou97}, raise questions about
what is required of a numerical calculation to obtain convergence in
the inviscid limit, or indeed what it means to converge.

\goodbreak
\section{Basic Equations}
\label{sec:equations}

We assume an isentropic body in uniform rotation at angular velocity 
$\vOmega$.
The linearized equations of motion in the corotating frame are
\begin{eqnarray}
  \label{eq:euler1}
  -i\omega\vv  +2\vOmega\cross\vv
   &=&-\grad\left(\cs^2\frac{p_1}{\rho}+\Phi_1\right)\equiv-\grad\psi\,,\\
   \label{eq:cont1}
   -i\omega\rho_1 +\grad\bdot(\rho\vv) &=&0.
\end{eqnarray}
Where necessary, first-order eulerian perturbations have been
distinguished by a subscript from the corresponding quantities in the
background state: for example, first-order mass density $\rho_1$ and
pressure $p_1$.  All such perturbations have the time dependence
$\exp(-i\omega t)$, $\omega$ being the angular frequency of the tide
viewed from the corotating frame.  Since the unperturbed fluid
velocity vanishes in this frame, $\vv$ is understood to be of first
order even though it is not explicitly marked as such.  We have
introduced $\psi$ as the sum of the first-order enthalpy
$p_1/\rho=\cs^2\rho_1/\rho$ and gravitational potential $\Phi_1$,
where $\cs^2\equiv(\partial p/\partial\rho)_S$ is square of the sound
speed.

The gravitational potential can be further subdivided as
$\Phi=\Phi_{1,\rm ext}+\Phi_{1,\rm self}$, the first term representing
the tidal potential exerted by the companion, and the second
representing the perturbation to the self-gravity of the fluid body.
For a fully self-consistent treatment, the equations above should be
supplemented by Poisson's equation $\nabla^2\Phi_{1,\rm self}=4\pi
G\rho_1$, but it will not be necessary to deal with $\Phi_{1,\rm
  self}$ explicitly in the present paper, since we are mainly
concerned with short-wavelength inertial modes for which self-gravity
is unimportant.  Even when we focus on the large-scale ellipsoidal
distortion of the body, for which the self-potential is important,
we pretend that it has already been computed and included in $\Phi_1$.

Equation~(\ref{eq:euler1}) can be solved algebraically for the velocity:
\begin{equation}\label{eq:vsol}
\vv =  \frac{1}{4\Omega^2-\omega^2}\left[
i\omega\grad\psi+2\boldsymbol{\Omega\times\nabla}\psi
+(i\omega)^{-1}4\boldsymbol{\Omega\Omega\cdot\nabla}\psi\right]
\equiv\mathbf{M}_\omega\bdot\grad\psi.
\end{equation}
This defines $\mathbf{M}_\omega$ as $3\times 3$ matrix or tensor that
is spatially constant in cartesian coordinates.  Using (\ref{eq:vsol})
to eliminate $\vv$ from the linearized continuity equation
(\ref{eq:cont1}), and remembering that $\psi=\Phi_1+\cs^2\rho_1/\rho$
leads to a wave equation for $\psi$:
\begin{eqnarray}
    \label{eq:wavey}
  \frac{4\Omega^2-\omega^2}{\cs^2}(\psi-\Phi_1)&=&\frac{1}{\rho}\grad\bdot
\left[\rho\left(\grad\psi+\frac{2}{i\omega}\boldsymbol{\Omega\times\nabla}\psi
-\frac{4\vOmega\vOmega}{\omega^2}\bdot\grad\psi\right)\right].
\end{eqnarray}
For application to hot Jupiters and binary stars, the tidal frequency
$\omega$ is typically comparable to the rotation frequency $\Omega$,
and both of these are small compared to the dynamical frequency
$\omega_{\rm dyn}\equiv (3G\Mp/\Rp^3)^{1/2}$.  Throughout most of the
planet, $\cs\sim\omega_{\rm dyn} \Rp$.  It follows that the lefthand
side of eq.~(\ref{eq:wavey}) is negligible throughout most of the interior
for disturbances whose wavelength $\lambda$ is small compared to the
planetary radius $\Rp$.  However, the lefthand side is important near
the surface of the planet; in fact it is singular there, because
$\cs^2\to 0$ at the surface, at least with idealized zero-temperature,
zero-pressure boundary conditions.  The righthand side of
eq.~(\ref{eq:wavey}) contains a term $\propto\vv\bdot\grad\ln\rho$
that is also singular at the surface.  In order that there be a
well-behaved solution for $\psi$ and $\vv$, it is necessary that these
two singular terms should balance: $\vv\bdot\grad\ln\rho\to
i\omega(\psi-\Phi_1)/\cs^2=i\omega\rho_1/\rho$.  Using the hydrostatic
equilibrium of the unperturbed state, this can be recast as
\begin{equation}
\label{eq:freebc}
  \vv\bdot\vn= \frac{i\omega}{g}(\Phi_1-\psi)
  \quad\mbox{at the surface, $R=\Rp$}\,,
\end{equation}
where $\vn$ is the outward-pointing normal to the unperturbed boundary
and $g>0$ is the effective gravity.  This is the usual free boundary
condition: it says that the lagrangian, not eulerian, enthalpy
perturbation vanishes at the surface.  The normal components of the
velocity and of the displacement $\boldsymbol{\xi}=\vv/(-i\omega)$ do
not vanish at the boundary.  It is true that it is possible to divide
the tidal response into long-wave and short-wave parts, $\vv=\vv_{\rm
  long}+\vv_{\rm short}$, in such a way that it would be an excellent
approximation to neglect $\vn\bdot\vv_{\rm short}$ at the boundary.
But then the analog of equation (\ref{eq:wavey}) for $\psi_{\rm
  short}$ would contain inhomogenous terms involving $\psi_{\rm long}$
even if the terms $\propto\cs^{-2}$ were neglected.  While the
principle of the free boundary condition is familiar, we have
emphasized the point because it will be important to our discussion of
the results of \cite{Wu05b} in \S\ref{subsec:Wu}.

For the time being,
we represent the rocky core by a rigid sphere of radius $\Rc<\Rp$:
\begin{equation}
  \label{eq:corebc}
  \boldsymbol{v\cdot\hat n}=0\quad\mbox{at $R=\Rc$,}
\end{equation}
where $\vn$ is now the normal to the core.  We take $\vn$ to point
away from the fluid, that is, downward at the core and upward at the
surface.  In \S\ref{subsec:rigidity}, we show that that the core
should deform with the equilibrium tide, and that it is more likely to
be liquid rather than solid.  To accomodate this deformation, the
tidal $Q$ derived for a completely rigid core [eq.~\eqref{eq:Qval}]
requires an overall correction factor of order unity that depends upon
the density contrast between the core and the convective region.

\subsection{Incompressible limit}\label{subsec:incomp}

Many of the calculations of this paper will be carried out in the
limit $\cs^2\to\infty$.  For a consistent hydrostatic equilibrium in the
unperturbed state, the background density must be constant.  In this
limit, eq.~(\ref{eq:wavey}) simplifies to
\begin{eqnarray}
  \label{eq:waveyinc}
  \nabla^2\psi - \left(2\vOmega\bdot\grad\right)^2\psi&=&0,\nonumber\\
\mbox{or}\quad \frac{\partial^2\psi}{\partial x^2}
+\frac{\partial^2\psi}{\partial y^2} -\frac{1}{c^2}
\frac{\partial^2\psi}{\partial z^2}&=&0.
\end{eqnarray}
in cartesian coordinates with the $z$ axis parallel to $\vOmega$.
We have introduced the abbreviation
\begin{equation}
  \label{eq:cdef}
  c\equiv\frac{\omega}{\sqrt{4\Omega^2-\omega^2}}
\end{equation}
for the coefficient of the vertical derivatives.  It is dimensionless,
unlike the sound speed $\cs$.  When $\omega^2<4\Omega^2$,
equation (\ref{eq:waveyinc}) is hyperbolic, $z$ plays
the role of a timelike coordinate,
and $c$ clearly plays the role of wave ``speed.''  The hyperbolic
nature of the more general equation (\ref{eq:wavey}) in the
low-frequency regime has been
noted by \cite{Savonije_Papaloizou97} and emphasized by OL04.

Equation (\ref{eq:waveyinc}) has been studied extensively for rotating
incompressible fluids \citep[e.g.][]{Greenspan}.  In the
incompressible limit, the potential perturbation does not enter the
wave equation at all.  But it does enter the free boundary condition
(\ref{eq:freebc}), which is still applicable.

A constant-density body is not a realistic model for a planet or star.
However, in addition to simplifying calculations, this model 
exhibits particularly clearly  the division of the tidal response between
short and long-wavelength parts, which is the main object of this paper.

\subsection{WKB dispersion relation and group velocity}\label{subsec:WKB}
For waves sufficiently short that they may be described locally
by plane waves proportional to $\exp(i\vk\bdot\vx-i\omega t)$, and for
frequencies $\omega^2\ll\cs^2 k^2$, equations (\ref{eq:wavey})
or (\ref{eq:waveyinc}) lead to the dispersion relation
\begin{equation}
  \label{eq:WKBdisp}
  \omega^2=\left(\frac{2\vOmega\bdot\vk}{k}\right)^2\,,
\end{equation}
where $k=|\vk|$.  For free oscillations, presuming that the components
of the wavevector $\vk$ are at least approximately real (so that the
envelope of the wave varies slowly with position), the dispersion
relation requires that $-2\Omega\le\omega\le 2\Omega$.  It is
important that $\omega$ is independent of wavelength $\lambda=k/2\pi$;
it depends only on the direction $\boldsymbol{\hat k}=\vk/k$, which
must lie on a double-napped cone whose axis is vertical, i.e. parallel
to $\vOmega$, and whose half angle is
\begin{equation}
  \label{eq:betadef}
  \beta\equiv\cos^{-1}\left|\frac{\omega}{2\Omega}\right|.
\end{equation}

Hereafter, to fix important signs, we adopt the conventions
$\omega\ge0$ and $\Omega>0$.  Tidal components that are retrograde
with respect to the planetary spin will be represented by negative
values of the azimuthal quantum number $m$ rather than negative values
of $\omega$: that is, nonaxisymmetric tides have the $(\phi,t)$
dependence $\exp(im\phi-i\omega t)$ with $m\ne0$, and their azimuthal
pattern speed is $\omega/m$.  With this convention,
$\omega=|2\vOmega\bdot\vk|/k$, and the group velocity becomes
\begin{equation}
  \label{eq:vgroup}
\vvg = \frac{\partial\omega}{\partial\vk}= 
\mbox{sign}(k_z)\frac{\vkh\cross(2\vOmega\cross\vkh)}{k}.
\end{equation}
So the waves and the energies they carry move crabwise, at right
angles to their wavevectors.

The time-averaged energy density and energy flux carried by inertial
waves are
\begin{mathletters}\label{eq:action}
  \begin{eqnarray}
    \label{eq:Edens}
    \mathcal{E} &=& \frac{1}{4}\rho\vv^*\bdot\vv = \frac{\rho k^2}
{2(4\Omega^2-\omega^2)}|\psi|^2,\\
    \label{eq:Eflux}
    \mathcal{F} &=& \mathcal{E}\vvg,
  \end{eqnarray}
\end{mathletters}
if the physical velocity perturbation is the real part of $\vv$.

\subsection{Tidal response of a coreless incompressible planet}
\label{subsec:corefree}
Inasmuch as the planet is small compared to its orbital semimajor axis
($a$)---typically $\Rp/a\sim 10^{-2}$ for hot Jupiters---the dominant
component of the external tidal potential is quadrupolar, that is,
proportional to $R^2 Y_{2,m}(\theta,\phi)\exp(-i\omega t)$ in
spherical polar coordinates $R\theta\phi$ centered on the planet.
Expressed in cartesians, $\Phi_{1,\rm ext}$ is then a linear
superposition of the terms
\begin{mathletters}  \label{eq:quadtide}
\begin{eqnarray}
  \label{eq:m2tide}
  m=\pm2: &\quad& A_{\pm 2}(x\pm iy)^2\,\\
  \label{eq:m1tide}
  m=\pm1: &\quad& A_{\pm 1}z(x\pm iy)\,\\
  \label{eq:m0tide}
  m= 0:  &\quad& A_0 (2z^2-x^2-y^2),
\end{eqnarray}
\end{mathletters}
times $\exp(-i\omega t)$, where the coefficients $A_{\pm2}$, $A_{\pm1}$
and $A_0$ are complex constants.
On the other hand, it is clear that the wave
equation (\ref{eq:wavey}) can be satisfied by polynomials in the cartesian
coordinates.  If the unperturbed body is axisymmetric---as it should be when
$\Omega^2\ll\omega_{\rm dyn}^2$)---then after transients have
died away, the forced responce of $\psi$ must have
the same azimuthal symmetry as that of $\Phi_{1,\rm ext}$.
If this response is also of second degree in the coordinates (an assumption
that turns out to yield an acceptable solution), then its components
at $m=\pm 1$ and at $m=\pm 2$ must have the same form as the first
two of equations (\ref{eq:quadtide});  it is easily seen that these
satisfy the wave equation (\ref{eq:waveyinc}).
But the axisymmetric component must have the form
\begin{eqnarray}
  \label{eq:m0resp}
m=0:\quad 
  \psi &\propto& 2c^2z^2+x^2+y^2~+\mbox{constant},
\end{eqnarray}
which differs from that of the perturbing potential,
eq.~(\ref{eq:m0tide}).  

Following eq.~(\ref{eq:vsol}), the displacements associated with these
quadratic forms of $\psi$ are linear functions of the coordinates.
From this and the constancy of the density in the interior, it is
easily shown that $\Phi_{1,\rm self}\propto\Phi_{1,\rm ext}$, so that
the total potential perturbation $\Phi_1=\Phi_{1,\rm ext} +\Phi_{1,\rm
  self}$ is also a superposition of the terms (\ref{eq:quadtide}),
with an appropriate rescaling of the coefficients.

It remains to relate the amplitudes of $\psi$ and $\Phi_1$ using
the surface boundary condition (\ref{eq:freebc}).  If we neglect
the influence of centrifugal force on the unperturbed state on the
grounds that $\Omega^2\ll\omega_{\rm dyn}^2=4\pi G\rho$, the unperturbed
boundary is spherical, $x^2+y^2+z^2=\Rp^2$, so that, with use of
eq.~(\ref{eq:vsol}),
\begin{equation}
  \label{eq:vnop}
\boldsymbol{v\cdot\hat n}= v_R =
\frac{1}{i\omega R}\left[z\pd_z-c^2(x\pd_x+y\pd_y)+\frac{2i\Omega c^2}
{\omega}(y\pd_x-x\pd_y)\right]\psi.
\end{equation}
The solutions for the azimuthal harmonics of $\psi$ then turn out to be
\begin{mathletters}  \label{eq:psisol}
\label{eq:psim2resp}
\begin{eqnarray}
  \psi&=&\frac{3\omega(4\Omega^2-\omega^2)}{8\pi G\rho(\omega-m\Omega)}\Phi_1
\quad\mbox{if } m\in\pm2,\\
\label{eq:psim1resp}
  \psi&=&\frac{3\omega^2(4\Omega^2-\omega^2)}{8\pi G\rho(\omega^2 -2\Omega^2 - m\Omega\omega)}\Phi_1
\quad\mbox{if } m\in\pm1,\\
\label{eq:psim0resp}
  \psi&=&\frac{3(\omega^2-4\Omega^2)}{8\pi G\rho}(x^2+y^2+2c^2z^2)
\frac{\Phi_1}{2 R^2 P_2(\cos\theta)}
\quad\mbox{if } m=0.
\end{eqnarray}
\end{mathletters}
These have been calculated to lowest order in $\omega^2/\pi G\rho$ and
$\Omega^2/\pi G\rho$, which are much less than unity.  At this level
of approximation, $\psi$ is negligible on the righthand side of the
surface boundary condition (\ref{eq:freebc}), which therefore reduces
to eq.~(\ref{eq:eqtidexi}), exactly as for the traditional
equilibrium tide when rotation is neglected.  Also to this order in
$\Omega^2/\pi G\rho$, the boundary is spherical.
A constant term in $\psi$ is not required even when $m=0$, since
the radial velocity and radial displacement
derived by substituting eq.~(\ref{eq:m0resp}) into
eq.~(\ref{eq:vnop}) turn out to be $\propto P_2(\cos\theta)$; this
is no miracle, but rather a consequence of the fact that
\begin{equation}
  \label{eq:eqtidexi}
\xi_R\approx -\frac{\Phi_1}{\bar g}\times[1+O(\Omega^2/\pi G\rho)]\,,
\end{equation}
where $\bar g$ is the angular average of the surface gravity
($=4\pi G\rho R_p/3$ for a constant-density planet).

Inasmuch as $\omega$ and therefore $c$ are real, the responses
(\ref{eq:psisol}) are perfectly in phase with the tidal forcing, so
that there is no secular input of tidal energy to this (highly
idealized) planet.  An exception might occur if the denominator in
eq.~(\ref{eq:psisol}) were to vanish, when the tide would be in
resonance with an $(\ell,m)=(2,1)$ free-precession mode.  But this
happens only at the single discrete frequency $\omega=\Omega$ [in
inertial space, $\omega=2\Omega$], and then only if the spin and orbit
are misaligned.  So it cannot be a general explanation for tidal
dissipation.  Also, the responses (\ref{eq:psisol}) are entirely
smooth and long-wavelength---or rather, \emph{nonwavelike}, since the
individual components of $\psi$ in (\ref{eq:psisol}) have no radial
nodes at $R>0$, as is the case for $\Phi_1$ itself.  Short-wavelength
inertial oscillations are not excited.  Note that the inner
boundary condition (\ref{eq:corebc}) is satisfied because we are
assuming that $\Rc=0$, and because all components of the velocities,
being linear functions of the coordinates, vanish at the origin.

These results are hardly new.  Solutions for the circular but not
necessarily synchronized case are known as Roche-Riemann ellipsoids
\citep{Chandra}.  Viewed in the corotating frame of the orbit rather
than that of the body's spin, they appear as ellipsoids with
stationary axes, though the flow velocity is generally nonzero in this
frame.  The axis ratios of these classical solutions are not limited
to values near unity, as ours are; they are exact nonlinear solutions.
Though exact, they are not always stable \citep{Chandra}.  In fact
\cite{Lebovitz_Lifschitz96} have shown that except for the synchronous
solutions, the Riemann and Roche-Riemann ellipsoids are generically
vulnerable to small-scale parametric instabilities, with growth rates
that increase with increasing departure from stationary and circular
streamlines.  However, the growth of small-scale inertial waves by
parametric instability is entirely distinct from direct tidal forcing,
and the secular energy dissipation rate that results is much more
difficult to estimate because it depends upon the amplitudes at which
these instabilities saturate, which necessarily involves nonlinear
considerations \citep[and references therein]{Arras_etal03}.

We conjecture that the absence of small-scale waves from the tidal
response is not a peculiarity of the incompressible, constant-density
limit, but that it holds more generally for isentropic, coreless,
compressible bodies whose unperturbed enthalpy profiles are
sufficiently smooth, provided that the outer boundary is free. This
conjecture, if true, would seem to contradict the results of
\cite{Wu05b}.  More will be said on this subject in \S\ref{subsec:Wu}.

\goodbreak
\section{Production of short waves by scattering from the core}
\label{sec:scattering}
When the radius of the rigid core is nonzero, it does not seem to be
possible to construct a purely nonwavelike tidal response, confirming
what OL04 concluded from numerical computations with a compressible
model. As far as we can tell, this remains true even in our
incompressible, constant-density model, at least in the limit that the
tide is strictly periodic. It seems that it ought to be possible to
prove (or disprove) this rigorously.  We have no such proof, but we
explain briefly why we think it is true in \S\ref{subsec:homogeneous};
the material of that subsection is not used in the rest of the paper.
Then in \S\ref{subsec:WKBscatt}, we go on to the main goal of this
section, which is to estimate the production of short waves at the
core on the \emph{assumption} that the response is partly wavelike.

\subsection{Homogeneous solutions of the wave equation}
\label{subsec:homogeneous}

The difficulty in constructing a purely nonwavelike response appears
to lie in the lack of a sufficiently rich set of homogenous functions
that solve equation (\ref{eq:wavey}) and are nonsingular on the
surface of a sphere, or indeed any closed surface.  A function
$f(x,y,z)$ is said to be homogenous of degree $D$ if $f(\sigma
x,\sigma y,\sigma z)=\sigma^D f(x,y,z)$ for any constant $\sigma\ne
0$; the degree may be negative. A homogeneous function cannot have
isolated radial nodes, since if it vanishes at any point, then it must
vanish at all points along the same radial ray.  But of course a
linear combination of such functions may have radial nodes.

If $c^2<0$, as when $\omega^2>4\Omega^2$ [cf.
eq.~(\ref{eq:cdef}], then equation (\ref{eq:wavey}) is elliptic
rather than hyperbolic, and after the rescaling $z= z/\sqrt{-c^2}$,
reduces to Laplace's equation.  Solid spherical
harmonics $r^\ell Y_{\ell,m}(\theta,\phi)$, when expressed in cartesians,
are homogeneous solutions---in fact polynomials---of degree $D=\ell\ge0$.
Each of these functions has a companion $r^{-\ell-1}Y_{\ell,m}(\theta,\phi)$
with negative degree, $D=-(\ell+1)$, that also solves Laplace's equation.
Given smooth values of $\psi$ or $\pd\psi/\pd n$ on concentric spheres
or (less conveniently) on spheroids, it is possible to construct a
smooth solution of Laplace's equation in the space between them as
linear combination of these functions; in general, unless the boundary
conditions are very restricted, both signs of $D$ are required.

As is well known, the spherical harmonics of a given degree $\ell$
constitute an irreducible representation of the rotation group
$SO(3)$, to which the laplacian is invariant.  This suggests that when
$\omega^2<4\Omega^2$, we should be concerned with homogeneous
functions that belong to irreducible representations of $SO(2,1)$, the
2+1-dimensional version of the Lorentz group, since that is the
group under which the d'Alembertian operator in eq.~(\ref{eq:wavey})
is invariant when $c^2>0$.  Let $s^2\equiv x^2+y^2-(cz)^2$ and
let $\eta\in(-\infty,\infty)$ be a real-valued hyperbolic angle,
and let $\phi\in(-\pi,\pi]$ be the usual azimuthal angle.
Then the following parametrizations cover the ``future,'' ``past'' and
``absolute elsewhere'' of the origin, respectively:
\begin{align*}
(x,y,cz)&=(|s|\cos\phi\sinh\eta,\,|s|\sin\phi\sinh\eta,\,|s|\cosh\eta)
 & s^2 &<0,~z>0;\\
(x,y,cz)&=(|s|\cos\phi\sinh\eta,\,|s|\sin\phi\sinh\eta,\,-|s|\cosh\eta)
& s^2 &<0,~z<0;\\
(x,y,cz)&=(|s|\cos\phi\cosh\eta,\,|s|\sin\phi\cosh\eta,\,|s|\sinh\eta)
 & s^2 &>0\,.
\end{align*}
Suitable homogeneous polynomials that satisfy eq.~(\ref{eq:wavey}) in
these three regions and belong to irreducible representations of
$SO(2,1)$ are
\begin{equation}\label{eq:wavepolys}
(-s^2)^{\ell/2}i^m Y_{\ell,m}(i\eta,\phi)\quad
(-s^2)^{\ell/2}i^m Y_{\ell,m}(\pi-i\eta,\phi),\quad\mbox{and}\quad
(s^2)^{\ell/2}i^m Y_{\ell,m}({\textstyle\frac{\pi}{2}}-i\eta,\phi).
\end{equation}
The tidal response of a coreless incompressible planet is made up of
linear superpositions of these: for example, 
eq.~(\ref{eq:psim0resp}) is proportional to $s^2 Y_{2,0}(i\eta,\phi)$ in
the ``future.''  In order to accommodate the boundary condition
$v_R=0$ at the surface of a finite core, one would like to add to the
polynomial solutions (\ref{eq:psisol}) a suitable linear combination
of homogenous solutions of negative degree. The
negative-degree solutions would be larger near the core than near the
surface, and therefore could ``patch up'' the boundary condition at the
core without much spoiling the boundary condition at the surface.
The obvious negative-degree counterpart to the first of the
functions eq.~(\ref{eq:wavepolys}), by analogy with the case of
Laplace's equation, is $(-s^2)^{-(\ell+1)/2}i^m
Y_{\ell,m}(i\eta,\phi)$.  While this is indeed a solution of the wave
equation, it is unfortunately singular where the core intersects the
``light cone'' $s^2=0$, i.e. at the critical colatitudes
$\theta=\beta$ and $\theta=\pi-\beta$ [eq.~(\ref{eq:betadef})].  So it
seems that we cannot superpose a finite number of finite-degree
homogeneous solutions to construct a smooth response.

\subsection{WKB scattering calculation}
\label{subsec:WKBscatt}
\subsubsection{Nonspecular reflection from a planar boundary}
\label{subsubsec:nonspecular}

To see how short inertial waves may be produced from long ones,
consider the reflection of an incident plane wave,
\begin{equation*}
\psi_{\rm in}(\vx,t)= A_{\rm in}e^{i\vk\bdot\vx-i\omega t}\,,
\end{equation*}
from a planar wall with normal $\vn$ pointing away from the fluid.
The reflection is generally not specular, and the incident and
scattered wavelengths differ, unless $\vn$ is parallel or
perpendicular to $\vOmega$.  These are consequences of the anisotropy
of the dispersion relation (\ref{eq:WKBdisp}) and would also hold,
with some quantitative changes, were buoyancy important.

If the wall is fixed in the unperturbed frame of the fluid and
sufficiently rigid so that any transmitted wave is negligible, then
energy conservation requires a reflected wave.  This is enforced by
the boundary condition, which we have taken to be $\vn\bdot\vv=0$, but
any other non-absorbing boundary condition would lead to the same
relationship between the incident and reflected wavevectors as the one
we are about to derive.

The scattered (outgoing) wavevector $\vk'$ is determined by two
conditions.  First, in order that the incident and scattered wave have
the same relative phase at all points along the wall, as required by
the boundary condition that connects them, it is necessary that $\vk$
and $\vk'$ have the same components parallel to the wall: that is,
$\vn\cross(\vk'-\vk)=0$, or equivalently
\begin{equation}
  \label{eq:samekwall}
  \vk'= k'_\perp\vn + (\vk-\vn\vn\bdot\vk)\equiv
  k'_\perp\vn + \vk_\parallel\,,
\end{equation}
where $k'_\perp$ remains to be determined.  Similarly, in order that
the relative phase be constant in time, the two waves must have
the same frequency, $\omega$.
Substituting from eq.~(\ref{eq:samekwall}) for $\vk'$ into
$(k')^2\omega^2=(2\vOmega\bdot\vk')^2$ leads to
\begin{equation}\label{eq:kquad}
  \left[(2\vOmega\bdot\vn)^2-\omega^2\right](k'_\perp)^2 
+8(\vOmega\bdot\vk_\parallel)
(\vOmega\bdot\vn)k'_\perp +(2\vOmega\bdot\vk_\parallel)^2
-\omega^2 k_\parallel^2=0.
\end{equation}
It is clear that one of the roots of this quadratic equation must be the
known solution $k'_\perp=k_\perp$ representing the incident wave, 
and therefore the product of the distinct roots is
\begin{equation}
  \label{eq:kroots}
  k_\perp k'_\perp = \frac{\omega^2\vk_\parallel^2
-(2\vOmega\bdot\vk_\parallel)^2}{\omega^2-(2\vOmega\bdot\vn)^2}\,.
\end{equation}
The denominator in eq.~(\ref{eq:kroots}) may vanish when the numerator
does not: the reflected wavevector then becomes infinite and normal to
the boundary.  We believe that the change of wavelength upon
  reflection, which is a direct consequence of the inertial-wave
  dispersion relation, underlies the singular behavior observed by
  OL04 in their numerical calculations.

\begin{figure}\epsscale{0.80}
  \plotone{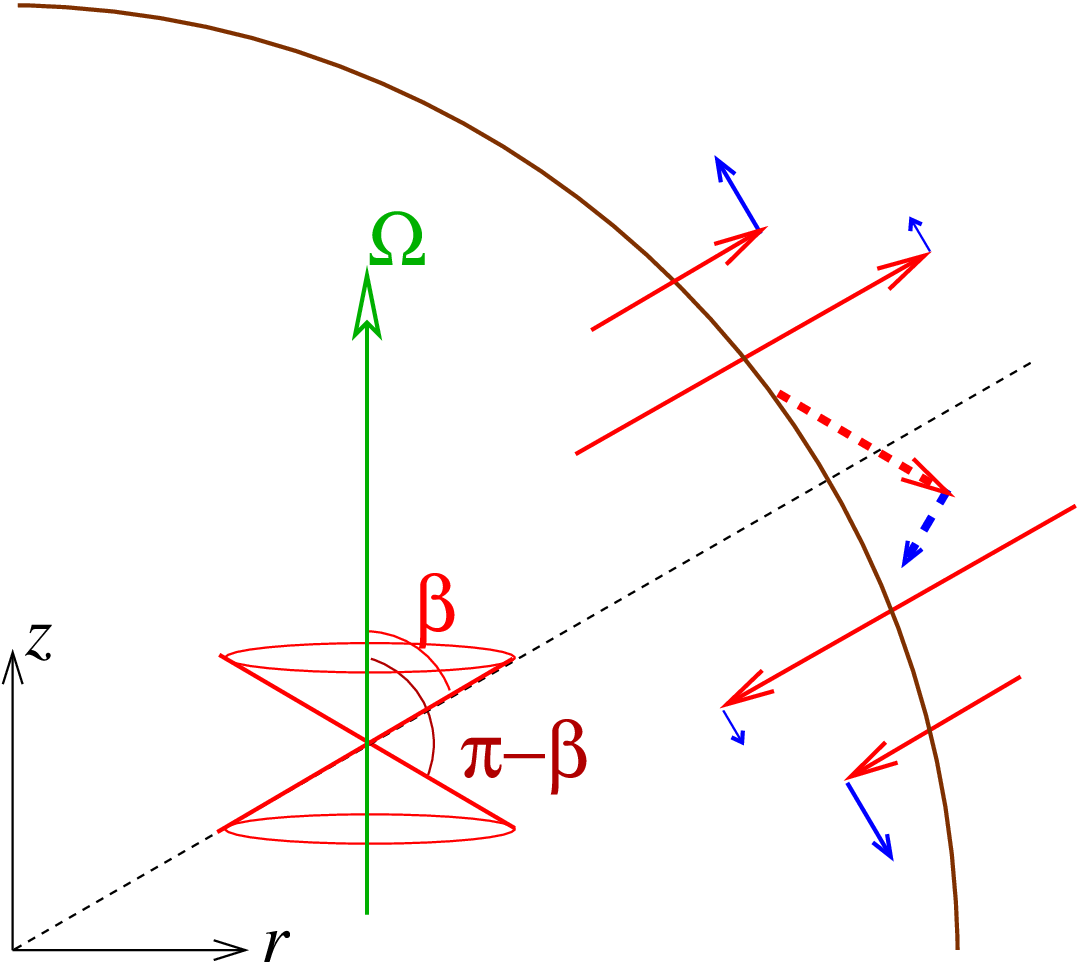}
  \caption{Scattering of a long wave into a short one at the core near
    the northern critical latitude. Allowed directions for $\vk$ make
    angles $\beta$ or $\pi-\beta$ with $\vOmega$, as shown by ``light
    cone.''  Large diagonal solid arrows (red in online version) are
    outgoing $\vk_{\rm out}$. These become infinite and change sign
    across critical latitude (thin dashed line). Smaller solid arrows
    (blue online) show group velocity $\boldsymbol{V}_{\rm g, out}$,
    whose magnitudes are $\propto k^{-1}$. Heavy dashed arrows show
    the incoming $\vk_{\rm in}$ and its associated group velocity.
    These are not to scale: $k_{\rm in}\ll k_{\rm out}$ and $V_{\rm g,
      in}\gg V_{\rm g, out}$.  In fact, all of the red arrows should
    have latitudinal components that are nearly the same in magnitude
    and sign, $\approx k_{\theta,\rm in}>0$.}
  \label{fig:fig1}
\end{figure}

\subsubsection{Scattering from a spherical core}\label{subsubsec:corescatt}

Now we apply this idea to scattering of the ``equilibrium tide''
from the spherical core.  As noted in \S\ref{subsec:corefree}, the
equilibrium tide is radially nodeless, and in this sense nonwavelike.
However, the functions (\ref{eq:psisol}) do have nodes in angular
directions: the wavenumber parallel to the surface of the core of
these quadratic functions is roughly
\begin{equation}
  \label{eq:kcore}
  \vk_\parallel \approx \pm\frac{2}{\Rc}\ve_\theta
\pm \frac{m}{\Rc\sin\theta}\ve_\phi\equiv k_\theta\ve_\theta+k_\phi\ve_\phi
\end{equation}
at colatitude $\theta$.  Since $|k_\parallel\Rc|>1$, we may expect
that WKB should be applicable to an outgoing wave whose radial
wavenumber $k'_\perp$ is real and $\gg 1/\Rc$.  We cannot make use of
eq.~(\ref{eq:kroots}) because the incident component $k_\perp$ is
ill-defined and the equilibrium tide does not satisfy the
WKB dispersion relation.  But equation (\ref{eq:kquad}) should still be
applicable to the short-wavelength outgoing wave.  It can be seen that
the coefficient of $k'_\perp$ vanishes at the critical latitudes
$\theta=\beta$ and $\theta=\pi-\beta$, where $\beta$ is defined by
eq.~(\ref{eq:betadef}).

The large root of eq.~(\ref{eq:kquad}) can be found by balancing
the terms in $k'_\perp$ and $(k'_\perp)^2$:
\begin{eqnarray}\label{eq:kRsol}
  k'_R &\approx& \frac{8(\vOmega\bdot\vk_\parallel)(\vOmega\bdot\vn)}
{\omega^2-(2\vOmega\bdot\vn)^2} = \frac{k_\theta\sin 2\theta}
{\cos^2\theta-\cos^2\beta}\nonumber\\[1ex]
&\approx& 
\begin{cases} 
k_\theta\csc(\beta-\theta) & \mbox{near}~\theta=\beta\\
k_\theta\csc(\beta+\theta) & \mbox{near}~\theta=\pi-\beta.
\end{cases}
\end{eqnarray}
Let us now determine when the large root represents an outgoing wave.
Since $|k'_R|\gg k_\parallel$, we may write 
$\vkh'\approx\mbox{sign}(k'_R)\ve_R+|k'_R|^{-1}\vk_\parallel$
through first order in $k_\parallel/k'_R$, whence from eq.~(\ref{eq:vgroup}),
\begin{equation}
  \label{eq:vgcrit}
  \vvg\approx 2\Omega\sin\beta\,\mbox{sign}(\cos\theta)\left(
\frac{k_\theta}{(k'_R)^2}\ve_R-\frac{1}{k'_R}\ve_\theta\right)
\end{equation}
near the critical latitudes.  The radial component of this is positive
when $\mbox{sign}(k_\theta)=\mbox{sign}(\cos\theta)$, so this becomes
the condition for the outgoing wave: $k_\theta>0$ at the northern
critical latitude, and $k_\theta<0$ at the southern one.  Also, it can
be seen from eqs.~(\ref{eq:vgcrit}) \& (\ref{eq:kRsol}) that the group
velocity is predominantly latitudinal and directed away from the
critical latitude: that is, towards the pole on the poleward side, and
toward the equator on the equatorial side.

Next, we determine the amplitude of the outgoing wave by matching it
to the positive-$k_\theta$ component of the incident wave via the
boundary condition $\ve_R\bdot(\vv_{\rm in}+\vv_{\rm out})=0$.
The radial velocity $v_R$ is related to $\psi$ by eq.~(\ref{eq:vnop})
in cartesians, and in spherical polars [cf. eq.~(\ref{eq:vsol})]
\begin{equation}
  \label{eq:vRpolar}
  v_R= \frac{1}{i\omega\sin^2\beta}\left[(\cos^2\theta-\cos^2\beta)
\frac{\pd\psi}{\pd R}-\frac{\sin2\theta}{2R}\frac{\pd\psi}{\pd\theta}
+\frac{m\cos\beta}{R}\psi\right].
\end{equation}
For the outgoing wave, this reduces at the critical latitudes to
\begin{equation*}
v_{R,\rm out}\approx \frac{|k_\theta| \Rc\sin\beta-im}
{2\Omega \Rc\sin^2\beta} \psi_{\rm out}\,.
\end{equation*}
The contribution from $\pd_R\psi$ is finite and
$\propto k_\theta$ despite the fact that $k'_R$ diverges, because of
the factor $(\cos^2\theta-\cos^2\beta)$ in front of $\partial\psi_{\rm
  out}/\partial R$.  For the ``incoming'' equilibrium tide, the
expression is similar, except that $\pd_R\psi_{\rm in}$ doesn't
contribute because $\partial\psi_{\rm in}/\partial R$ is finite at the
critical latitude:
\begin{equation*}
v_{R,\rm in}\approx \frac{-|k_\theta| \Rc\sin\beta-im}
{2\Omega \Rc\sin^2\beta} \psi_{\rm in}\,.
\end{equation*}
Combining the last two equations, we have the following relation
between the long-wavelength ``ingoing'' tide and the short-wavelength
outgoing wave:
\begin{equation}
  \label{eq:Ainout}
  A_{\rm out}\approx \frac{|k_\theta| \Rc\sin\beta- im}
{|k_\theta| \Rc\sin\beta+im}A_{\rm in}\,.
\end{equation}
Thus, the boundary condition $\xi_{R,\rm out}=-\xi_{R,\rm in}$
at the core leads to $|A_{\rm out}|=|A_{\rm in}|$.

The relationship between the amplitude $A_{\rm in}$ of the
``incident'' component of the equilibrium tide and the tidal potential
$\Phi_1$ is given by equations (\ref{eq:psisol}).  For definiteness,
consider a synchronization tide exerted on a planet in a circular
orbit, so that 
\begin{equation}
  \label{eq:freqs}
m=2\mbox{\,sign}(n-\Omega),\ \mbox{and}\ \omega=2|n-\Omega|,\ 
\mbox{where}\ n\equiv\frac{2\pi}{P_{\rm orb}}
\end{equation}
is the mean motion.  In this case,
the relevant component of $\Phi_1$ becomes
\begin{equation}
  \label{eq:synctide}
  \Phi_1\to -\frac{3fGM_*}{4a^3}R^2\sin^2\theta e^{im\phi-i\omega t}\,.
\end{equation}
The factor $f$ represents the ratio $\Phi_1/\Phi_{1,\rm ext}$ of the
total perturbing potential to that part which is exerted by the
companion.  In general, $f\approx1+2k_{\rm p}$ at the surface, $R=R_p$, where
$k_{\rm p}$ is the apsidal-motion constant, equal to
one half the planetary Love number.
For a constant-density coreless planet, $f=5/2$.
Expanding $\sin^2\theta$ and extracting the coefficient of
$\exp(2i\theta)$ leads to
\begin{equation}
  \label{eq:Async}
\quad A_{\rm in}= -\frac{9f}{16}\frac{\omega(\omega+m\Omega)}{8\pi G\rho}
\frac{GM_*R_c^2}{a^3}
\quad\mbox{for $|m|=2$ and $e=0$}.
\end{equation}

\subsubsection{Energy flux and power of the scattered waves}
\label{subsubsec:scatteredpower}

Since $|\psi_{\rm out}|=|A_{\rm in}|$ 
equations
(\ref{eq:action}), (\ref{eq:vgcrit}), and (\ref{eq:Async}) imply
that the radial component of the energy flux in the outgoing
short waves at the core is
\begin{equation}
  \label{eq:Fout}
  \mathcal{F}_R= \frac{\rho|A_{\rm in}|^2}{\Rc\sqrt{4\Omega^2-\omega^2}}
\approx 
\left(\frac{3f M_*\Rp^3}{32\Mp a^3}\right)^2\frac{\omega^2(\omega+m\Omega)^2}
{\sqrt{4\Omega^2-\omega^2}}\rho \Rc^3
\end{equation}
for the $|m|=2$ synchronization tide, i.e. for the tidal response of a
planet in a circular orbit with aligned but nonsynchronous rotation.
Expressions for $\omega$ and $m$ in terms of the orbital and
rotational periods were given above in \S\ref{subsubsec:corescatt}.
As it stands, eq.~(\ref{eq:Fout}) is valid only for a constant-density
planet (so that $f=5/2$), and only for the waves launched sufficiently
close to the critical latitude so that we may take
$|\boldsymbol{k}'|\approx|k'_R|\gg k_\theta,\,k_\phi$.  We have taken
$|k_\theta|\approx 2/\Rc$ and $k_\phi=m/R$, as appropriate for
the response (\ref{eq:psim2resp}) to an $|m|=\ell=2$ tidal potential.
Notice that the large radial wavenumber has canceled between the wave
energy density (\ref{eq:Edens}) and radial group velocity
(\ref{eq:vgcrit}), so that the flux is approximately independent of
latitude, provided that $\theta$ is sufficiently close to 
$\beta=\cos^{-1}(\omega/2\Omega)$ or to $\pi-\beta$.  Let us therefore
integrate over latitudinal bands of width $2\Delta\theta$
centered on both critical latitudes, 
with $\Delta\theta\ll\beta,\pi-\beta$, to obtain
the total mechanical power carried radially outward
by the short waves launched within these bands:
\begin{eqnarray}
  \label{eq:power0}
  \dot E(\Delta\theta)&\approx& 8\pi \Rc^2\sin\beta\Delta\theta\times\mathcal{F}_R
\nonumber\\
 &\approx & 2\pi\left(\frac{3f}{16}\right)^2\omega(\omega+m\Omega)^2
\left(\frac{M_*\Rp^3}{\Mp a^3}\right)^2\rho \Rc^5\Delta\theta.
\end{eqnarray}

Although eq.~(\ref{eq:power0}) has been derived for an incompressible,
constant-density planet, it can be generalized to an isentropic
compressible body with standard approximations for the equilibrium
tide.  In our approach where the core is regarded as a perturbation to
an otherwise homogeneous body, we continue to calculate the
equilibrium tide as if the core were absent.  What is needed for the
scattering calculation is the radial displacement of the equilibrium
tide at the core,\footnote{It is assumed here that $\xi_R$ and
  $\Phi_{1}$ are proportional to the same function of
  $(t,\theta,\phi)$ as the perturbing quadrupolar potential
  $\Phi_{1,\rm ext}$ of the star, so this dependence will be taken as
  read} $\xi_{R}^{\rm eq}(\Rc)$ For a constant-density body, the
radial strain $\sigma\equiv\xi_R(R)/R$ is independent of radius $R$,
so
\begin{equation}\label{eq:xireq}
  \xi^{\rm eq}_R(\Rc)=\frac{\Rc}{\Rp}\xi^{\rm eq}_R(\Rp)\approx
 -\frac{\Rc}{\Rp}\frac{\Phi_1(\Rp)}
  {g_{\rm p}}= -f\left(\frac{\Rp}{a}\right)^3\frac{M_*}{\Mp}\Rc,
\end{equation}
where $a$ is the semimajor axis of the orbit, $g_{\rm p}=G\Mp/\Rp^2$,
and the surface displacment has been evaluated from (\ref{eq:freebc})
with neglect of $\psi$ on the righthand side as before, i.e.
$\xi_R(\Rp)\approx-\Phi_1(\Rp)/g_{\rm p}$.  We assume that the
relation (\ref{eq:xireq}) between the surface displacement of the
equilibrium tide and the disturbing potential holds in general,
provided that $f$ is interpreted as $1+2k_{\rm p}$ for the relevant
value of the apsidal motion constant $k_{\rm p}$.  The latter is
$k_{\rm p}\approx0.260$ for an $n=1$ Emden polytrope, which roughly
approximates a Jovian planet, as compared to $k_{\rm p}=3/4$ for a
constant-density body ($n=0$ polytrope).  However, since the radial
strain of the tide is not constant with radius in general, we must
multiply $f$ in eq.~(\ref{eq:xireq}) by a factor
$\sigma(\Rc)/\sigma(\Rp)$ to obtain
the correct displacement at the core.  If the core is sufficiently
small, $\Rc\lesssim\Rp/2$, then $\sigma(\Rc)\approx\sigma(0)$.  

It is not immediately clear how to estimate $\sigma(0)$ easily.  One
possibility is to assume that $\xi_R^{\rm eq}$, which represents the
radial displacement of the fluid by the equilibrium tide, is the same
as the radial displacement of the equipotential surfaces between the
undistorted and distorted states; then it can be shown by integration
of the Radau equation from which $k_{\rm p}$ is obtained
\citep{Schwarzschild58} that $\sigma(0)/\sigma(\Rp)\approx(1+2k_{\rm
  p})^{-1}\approx 0.658$ for the $N=1$ polytrope, so that the factor
in square brackets in (\ref{eq:xireq}) reduces to unity for $N=1$.
Alternatively, if one evaluates $\xi_R^{\rm eq}$ from the summed
zero-frequency response of all the normal modes, taking into account
their overlap with the quadrupolar perturbing potential $\Phi_{1,\rm
  ext}$, then it can be shown that $\sigma(0)/\sigma(\Rp)\approx
0.813$ at $N=1$.  By far the largest response is that of the $\ell=2$
fundamental mode, for which $\xi_R$ is nearly linear in $R$.  These
two estimates of $\sigma(0)/\sigma(\Rp)$ differ because, even for an
originally spherical and nonrotating body, the radial displacements of
the fluid and of the equipotentials need not coincide except at the
surface.  There are many possible displacement fields that can be
compatible with a given distortion of the density and potential
fields, and it is not possible to choose among them on the basis of
the continuity equation alone without some auxiliary constraint.  It
can be shown that in a stratified region, the appropriate constraint
is $\boldsymbol{\nabla\cdot\xi}=0$ even for a compressible body: the
density and pressure of fluid elements is not disturbed by the
equilibrium tide in the stratified case.  OL04 adopted this
constraint.  For an isentropic region in a nonrotating star, the
appropriate constraint is $\boldsymbol{\nabla\times\xi}=0$ rather than
$\boldsymbol{\nabla\cdot\xi}=0$ since vorticity is conserved; in a
compressible body, this leads to a different pattern of fluid
displacements for the same distortion of the density and potential
\citep{Goodman_Dickson98,Terquem_etal08}.  Thus our second estimate
(the one yielding $0.813$) was calculated under the assumption that
$\boldsymbol{\xi}=\boldsymbol{\nabla}\chi$ for some scalar function
$\chi$.  Since our bodies rotate, however, the axial component of
vorticity does not vanish and therefore $\boldsymbol{\nabla
  \times\xi}=0$ may not be the correct constraint.\footnote{In the
  strict low-frequency limit where $\omega^2$ is small compared with
  $4\Omega^2$, and not just small compared to $\omega^2_{\rm dyn}$,
  the vorticity would remain axial under the equilibrium tide by the
  Taylor-Proudman theorem, so that the poloidal components of
  $\boldsymbol{\xi}$ would be curl-free.}  For the purpose of
estimating the order of magnitude of the tidal $Q$ that results from
scattering by the core, the differences among $0.658$, $0.813$, and
unity (the correct result for $N=0$) are not important, and we take
this as an indication that the true value of $\sigma(0)/\sigma(\Rp)$
is also sufficiently close to unity.

The frequency-dependent factor $\omega(\omega+m\Omega)$ in
eq.~(\ref{eq:power0}) becomes $8|n-\Omega|n^2$ with use of
(\ref{eq:freqs}).  Finally, the relevant density is the density of the
fluid at the surface of the core, $\rho(\Rc)$.  With these
modifications, the wave power (\ref{eq:power0}) generalizes to
\begin{equation}
\label{eq:power}
\dot E(\Delta\theta)\approx |n-\Omega|n^2
\frac{9\pi}{16}\left[(1+2k_{\rm p})\frac{\sigma(\Rc)}
  {\sigma(\Rp)}\right]^2\left(\frac{M_*}{\Mp}\right)^2
\left(\frac{\Rp}{a}\right)^6\rho(\Rc)\Rc^5\Delta\theta.
\end{equation}
The tidal torque, or more precisely, the rate of increase of the
angular momentum carried by the waves, is $\Gamma=\dot E\times
m/\omega= \dot E/(n-\Omega)$.  Thus according to eq.~(\ref{eq:power}),
the torque has the same sign as $n-\Omega$---meaning that
subsynchronous planets spin up and supersynchronous ones spin
down---but is independent of the magnitude of the departure from
synchronous rotation.  This is probably not true if the departure is
very small, $|n-\Omega|\ll n$: in that limit, the critical latitudes
converge upon the equator from both sides, whereas the approximations
used to derive eq.~(\ref{eq:power}) implicitly assume that these
latitudes are well separated from one another.

\subsubsection{Dissipation of the short waves, and the tidal $Q$}
\label{subsubsec:dissip}

Waves that do not
dissipate appreciably before returning to the region from which they
are launched (perhaps after multiple reflections between the outer
boundary and the core, with changes of wavelength at each reflection)
must be treated as global normal modes.  Secular input of energy and
angular momentum to nondissipative global modes would occur only at
exact resonance with the tide, which would almost never occur for
modes of finite wavelength.  Furthermore, 
the angular momentum carried by the waves is not transferred to
the mean flow, and therefore does not alter the angular velocity
$\Omega$, until those waves dissipate.  

Thus, in order to estimate $Q$, we must consider the dissipation of
the waves.  Unfortunately, this is a complicated issue.  More than one
process may be important, depending upon aspects of the planetary
structure and transport processes that can justifiably be neglected in
calculating the wave excitation.  Nevertheless, some dissipative
processes can be ruled out, and a rough upper limit on $Q$ as a function of
the core radius can be obtained.

First of all, the narrower the width $\Delta\theta$ of the latitudinal
bands around critical latitudes that we consider, the shorter is the
wavelength of the waves launched within those bands: eq.~(\ref{eq:kRsol})
shows that $\lambda\sim\pi\Rc\Delta\theta$.  All of the obvious
dissipation mechanisms become more efficient as wavelength decreases.
The rate of viscous dissipation, for example, scales as $\nu k^2$, and
therefore $\propto|\Delta\theta|^{-2}$.  Furthermore, the propagation
time from the core to the surface of the planet---or rather, to the
upper boundary of the convection zone, since in a realistic hot
Jupiter there must be a stratified region near the surface---scales
$\propto\lambda\propto\Delta\theta$.  [Eq.~(\ref{eq:vgcrit}) says that
the radial component of the group velocity starts out at $V_{{\rm gr},
  R}\sim \Omega\lambda^2/\Rc$.  But this is because the rays emanating
from near the critical latitude are almost tangential to the core.
The rays follow straight lines in the meridional plane, and $\lambda$
is approximately constant along them, so that a more representative
value of the group velocity for the purpose of estimating the
propagation time is $|\vvg|\sim
\Omega\lambda\sim\Omega\Rc\Delta\theta$.]  Hence the number of viscous
dissipation times per transit time scales $\propto |\Delta\theta|^{-3}$.

Nevertheless viscous dissipation is probably negligible, as can be
seen by very rough order of magnitude considerations.
\cite{Goldreich_Nicholson77} estimated that turbulent convective
viscosity acting on the equilibrium tide, which has a ``wavelength''
$\sim\Rp$, would yield $Q\approx 10^{13}$ because of suppression of
turbulent viscosity by a factor $(\omega\tau_{\rm c})^{-2}$, where
$\tau_{\rm c}$ is the turnover time of the largest eddies.  (When this
argument applies, the $Q$ based on the laminar viscosity should be
even larger.)  In our case, the suppression factor would not be quite
so small because the tidal period is a few days even for a
substantially nonsynchronous rotation, rather than 5 hours as for the
Jupiter-Io system.  So by Goldreich and Nicholson's reasoning, $Q\sim
10^{11}$ for the equilibrium tide in our case.  The short waves have
the same period as the tide itself, so the suppression factor is the
same for them, but their damping rate is increased by a factor $\sim
(kR)^2\sim|\Delta\theta|^{-2}$, and allowing for their transit time
between the core and the surface, we may conclude that the $Q$ value
due to turbulent convective damping of short waves should be at least
$10^{11}|\Delta\theta|^3$ in hot Jupiters.  We have stated this as an
inequality because the core is small and therefore somewhat
inefficient at scattering the equilibrium tide; this will be made more
quantitative below, but for now, note simply that the wave power
(\ref{eq:power}) is $\propto\Rc^5$.  To be astrophysically relevant,
$Q$ should be of order $10^5$ to $10^6$.  Therefore only waves within
$\Delta\theta\lesssim 10^{-2}$~radians of the critical latitude could
damp effectively by this mechanism.  But as we will soon show, other
mechanisms exist that can damp the waves launched farther from the
critical latitude, and since the wave power $(\ref{eq:power})$ is
proportional to $\Delta\theta$, these mechanisms give a smaller $Q$.

An important source of dissipation for short inertial waves in hot
Jupiters is escape from the convection zone, where most of the mass of
the planet resides, into the stratified radiative zone near the
surface, whose existence is guaranteed by strong illumination from the
host star.  In the radiative zone the waves convert into g~modes (more
properly, Hough modes) that are supported primarily by buoyancy rather
than Coriolis forces.  They are then subject to damping by radiative
diffusion because they perturb the temperature and entropy profiles.
Radiative diffusion damps more efficiently than viscosity because of
the much longer mean free path of photons compared to molecules or
ions, and it is all the more efficient because the wavelength in the
radiative zone shortens by a further factor $\sim\Omega/N\sim
10^{-2}$, where $N$ is the Brunt-V\"as\"al\"a frequency there.  The
upshot is that an outgoing short inertial wave that penetrates the
radiative zone will almost certainly damp before returning to the
core.  Details will be given by Eric Johnson in a forthcoming paper.
Here we simply note that penetration is not possible between the poles
and the critical latitudes, because the Hough modes are evanescent
there.  Using the fact that the group velocities lie at angles
$(\pi/2)\pm\beta$ with respect to the polar axis in the meridional
plane, one can show with a little trigonometry that penetration into
the radiative zone can occur at the first encounter only if
\begin{eqnarray}
  \label{eq:escape}
  \frac{\Rc}{R_{\rm s}}\cos\Delta\theta &<& \frac{\omega^2}{2\Omega^2}-1\quad
\mbox{(poleward rays)}\nonumber\\
  \frac{\Rc}{R_{\rm s}}\cos\Delta\theta &>& 1-\frac{\omega^2}{2\Omega^2}\quad
\mbox{(equatorward rays)},
\end{eqnarray}
where $R_{\rm s}\approx\Rp>\Rc$ is the radius at the convective-radiative interface.
A poleward ray is one that is launched between the critical
latitude and the pole, so that it starts out toward the
rotation axis, etc.  Neither inequality in (\ref{eq:escape})
can be satisfied when $\omega<\Omega\sqrt{2}$, i.e. $\beta>\pi/4$.  In
such cases the waves are fully reflected at their first encounter with
the radiative zone and return toward the core.  But the ray may enter
the radiative zone on a subsequent encounter.

Paradoxically, the easiest damping mechanisms to predict with
confidence by analytic means may be the nonlinear ones.  It is
reasonable to assume that any inertial wave whose velocity amplitude
satisfies $v\gtrsim\omega/k$ will quickly damp by some combination of
Kelvin-Helmholtz instabilities or three-mode coupling to even
shorter-wavelength daughter modes, as is observed for closely related
internal waves (g-modes), both in the laboratory and in the oceans
\citep[e.g.][]{McEwan71,Mueller_etal86}.  This process is effectively
local, occuring on lengthscales comparable to the wavelength and
timescales comparable to the wave period.  The dimensionless
nonlinearity parameter $kv/\omega$ diverges rapidly
toward the critical latitude.
From eqs.~(\ref{eq:action}), $|\vv|^2=4\mathcal{F}_R/\rho V_{{\rm g},R}$;
substituting then from eqs.~(\ref{eq:vgcrit}) and (\ref{eq:Fout}), and
generalizing the latter in the same way that we turned eq.~(\ref{eq:power0})
into eq.~(\ref{eq:power}), we find that
\begin{equation}
  \label{eq:nonlinearity}
  \frac{k v}{\omega}\approx\left[\frac{3}{2 \sqrt{8}} (1+2k_{\rm p})
    \frac{\sigma(\Rc)} {\sigma(\Rp)}\,
    \sqrt{\frac{\omega+m\Omega}{\omega-m\Omega}}\right]
    \frac{M_*\Rp^3}{\Mp a^3}\,(\Delta\theta)^{-2}\,.
\end{equation}
The contents of the square brackets above are close to unity.  Therefore,
nonlinear dissipation will dominate within latitudinal bands of halfwidth
\begin{equation}
  \label{eq:nonlinlat}
  \Delta\theta_{\rm nl}\approx \left(\frac{M_*\Rp^3}{\Mp a^3}\right)^{1/2}
\approx 0.03\left(\frac{\Rp}{R_{\rm J}}\right)^{3/2}
\left(\frac{\Mp}{M_{\rm J}}\right)^{1/2}\left(\frac{4\,{\rm d}}{P}\right)
\,\mbox{radians.}
\end{equation}
Though small compared to unity, this is indeed larger than the
generous estimate made above for the latitudinal distance within which
turbulent convective viscosity might be important.  We regard this as
a lower bound on the value of $\Delta\theta$ within which the waves are
able to dissipate, since other mechanisms---especially escape into the
radiative zone---may contribute.

The tidal $Q\equiv\omega\Delta E/\dot E$, where $\Delta E$ is the
maximum potential energy associated with the time-variable part of the
tidal distortion.  The instantaneous gravitational energy of the
equilibrium response to an applied quadrupole $\Phi_{1,\rm
  ext}(R,\theta,\phi)=K R^2 P_2(\cos\gamma)$ is $k_{\rm p}
K^2\Rp^5/G$; here $K$ is a constant and $\gamma$ is the angle between
the symmetry axis of the potential and the point $\boldsymbol{R}$.
For the present case of a synchronization tide, $K=-\sqrt{3/2}GM_*/a^3$
[eq.~\eqref{eq:ecntidefull}], so 
\begin{equation}
  \label{eq:deltaEsyn}
  \Delta E= \frac{3}{2}k_{\rm p}\frac{GM_*^2\Rp^5}{a^6}\,.
\end{equation}
 Using eq.~(\ref{eq:freqs}) \& (\ref{eq:power}) and
taking $k_{\rm p}\approx 0.26$ and $\rho(\Rc)\approx (\pi\Mp/4\Rp^3)$
as appropriate for the apsidal motion constant and central density of
a homogeneous $N=1$ polytrope, we have
\begin{eqnarray}
  \label{eq:Qval}
  Q&\approx& \frac{16k_{\rm p}}{3\pi^4}\left[(1+2k_{\rm p})\frac{\sigma(\Rc)}
  {\sigma(\Rp)}\right]^{-2} P^2\frac{G\Mp\Rp^2}{\Rc^5}(\Delta\theta)^{-1}
\nonumber\\[1ex]
&\approx& 2.1\times 10^7\left(\frac{P}{4\,{\rm d}}\right)^2
\left(\frac{\Mp}{\MJ}\right)\left(\frac{\Rp}{\RJ}\right)^2
\left(\frac{\Rc}{0.2\RJ}\right)^{-5}
\left(\frac{\Delta\theta}{0.1\,{\rm rad}}\right)^{-1}.
\end{eqnarray}
In the final line, we have taken the square brackets equal to unity.
This result is some two orders of magnitude larger than is typically
assumed for hot Jupiters but is obviously very sensitive to the
assumed core radius.  The fiducial value $\Rc\approx0.2\RJ$ is a crude
estimate based on the assumptions that $\Mc\approx0.1\Mp$ and that the
core must be roughly twice as dense as its immediate surroundings at
the same pressure because it has roughly twice the molecular weight
per electron.  Here $P$ is the period of the orbit ($P=2\pi/n$),
not the period of the tide ($P_{\rm tide}=\pi/|n-\Omega|$): $Q$
turns out to be independent of the latter.  The $Q$ for circularization
of a slightly eccentric but synchronous orbit is nearly the same
for this mechanism if the same $\Delta\theta$ applies (see the Appendix).

Values for $Q$ quoted in the literature are often normalized by the
tidal energy of a homogeneous body of the same mass and radius; see
\cite{Mardling07} \& \cite{FJG07} for discussion of this point.  Since
$k_{\rm p}=3/4$ rather than $\approx 0.26$ for a homogeneous body, the
values \eqref{eq:Qval} and \eqref{eq:Qecc} should be roughly tripled
to conform with that convention.

\section{Discussion}\label{sec:discussion}

Up to this point, we have had little to say about the work of Yanquin
Wu on the excitation of inertial waves in \emph{coreless} but
compressible isentropic fluid bodies \citep{Wu05a,Wu05b}.  Here we
explain why we believe that her calculations, though technically
superb, are not applicable to tides in real planets or stars.  Some of
these criticisms may apply also to the work of
\cite{Ivanov_Papaloizou07}, who adopted a ``low-frequency''
approximation that appears to be physically equivalent to Wu's.  But
we focus on Wu's work because of the admirable clarity of her
exposition, and because she is concerned with resonant excitation
rather than the quasi-impulsive excitation analyzed by
\cite{Ivanov_Papaloizou07}.

We then go on to discuss whether the rock-and-ice core of a jovian
planet can be regarded as rigid, or even solid, and the consequences
for the production of inertial waves if it is not.

\subsection{Previous tidal calculations for coreless isentropic bodies}
\label{subsec:Wu}

The salient claims by Wu that we address are the following:
\begin{enumerate}
\item The overlap integrals between the short-wavelength inertial modes and the
perturbing tidal potential diminish as a negative power of
the number of radial or latitudinal nodes, rather than exponentially,
even for unperturbed radial density profiles that are smooth apart from
a power-law convergence to zero at the surface.
\item For many radial nodes, the overlap is concentrated toward the
  surface; this is where the excitation mainly occurs.
\item The overlap integrals vanish for a constant-density, incompressible body.
\end{enumerate}
Wu suggested that the tidal coupling could be enhanced by a density
discontinuity associated with a first-order phase transition within
the convection zone, perhaps at the interface between molecular and
metallic hydrogen.  This is probably true in principle, but we address
here only the claims made for strictly isentropic bodies, thereby
excluding first-order phase transitions because of the entropy jump
associated with latent heat.  There does not appear to be a consensus as to
the first-order nature of the molecular-to-metallic transition in hydrogen
\citep[][and references therein]{Militzer_etal08,Sumi_Sekino08}.

We agree in part with the third of the ennumerated claims above but take issue with the
first two.  Our counter-arguments are mainly these:
\begin{enumerate}
\item By selectively neglecting compressibility in some places while
  retaining it in others, and especially by oversimplifying the
  low-frequency limit of the outer boundary condition, Wu has created
  a singularity at the boundary, though the dynamics there should
  actually be smooth if the boundary is free.  We suspect that the
  tidal coupling she calculates is due mainly or entirely to this
  singularity.
\item For barytropic bodies [$P=P(\rho)$], albeit with an inconsistent
  treatment of their gravitational potentials, there exist tidal
  responses that completely lack short-wavelength
  components and that can be exhibited in closed form.  The fluid
  displacement and enthalpy are independent of adiabatic index, and so
  are the same for a compressible as for an incompressible body.
\end{enumerate}
We now expand upon these last two points.

After deriving the equivalent of equation (\ref{eq:wavey}), Wu argues
that the lefthand side, which contains the sound speed $\cs$ in the
denominator, can be neglected on the grounds that compression of the
fluid is very slight for modes that are both short in wavelength
($kR\gg 1$) and low in frequency ($\omega^2\ll\cs^2k^2$).  This term
comes directly from the time derivative in the continuity equation, so
dropping it is equivalent to replacing the continuity equation by
$\diver(\rho\vv_1)=0$.  But according to her formalism, the tidal
forcing is explicitly $\propto\cs^{-2}$, as shown by eq.~(7) of
\cite{Wu05b}, so the tidal coupling itself should vanish in the limit
$\cs^{-2}\to0$.  Indeed, her Appendix~B makes this explicit.  However
the neglected term is singular at the free surface, because
$\cs^2\to0$ there. In the full equation (\ref{eq:wavey}), the
singularity involving $\cs^{-2}$ is balanced by another singularity
involving $\propto\rho^{-1}\grad\rho$, where $\rho$ is the unperturbed
density, Using our equations (\ref{eq:vsol}) \& (\ref{eq:wavey}), it
can be seen that the condition under which the two singularities
cancel one another is precisely the free boundary condition
(\ref{eq:freebc}), whose physical interpretation is the vanishing of
the lagrangian enthalpy perturbation.

If, following \cite[\S2.1 \& \S4.1]{Wu05a}, one neglects the righthand
side of eq.~(\ref{eq:wavey}) but retains the gradient of the
unperturbed density profile on the righthand side, then the only
possible well-behaved modes of the system are those for which the
normal components of the fluid velocity and displacement
[$\vxi=\vv/(-i\omega)$] vanish at the boundary.  Evidently, Wu
believed that taking $\boldsymbol{\hat n\cdot\xi}=0$ at the boundary
is an acceptable approximation because (i) the inertial modes don't
move the boundary very far, and (ii) the density vanishes at the
boundary anyway.  The modification to the boundary condition affects
the structure of the mode not just at the surface, however, but down
to depths comparable to that of the first radial node of $\psi$ below
the surface. Let $h$ be half the nodal depth, so that
$\psi(h)\approx\psi(0)$, where $\psi(0)$ is the surface value of the
eigenfunction.  Denote the unperturbed density at this depth by
$\rho(h)$, and the sound speed by $\cs(h)= \sqrt{gh(N+1)/N}$, assuming
a polytropic equation of state $P\propto\rho^{(N+1)/N}$.  The eulerian
density perturbation at this depth is
\begin{equation}
  \label{eq:eulerianpert}
  \rho_1(h)=\rho(h)\frac{\psi(h)}{\cs^2(h)}\approx\frac{N\rho(h)}{N+1}
\frac{\psi(0)}{gh}\,.
\end{equation}
(With Wu's definition of $\psi$, there would be a factor of $\omega^2$ here.)
On the other hand, since $\rho(R)\propto(\Rp-R)^N$ near the surface,
the contribution to $\rho_1=-\diver(\rho\vxi)$ from the surface
displacement, which Wu neglects (as do Ivanov \& Papaloizou), is
\begin{equation}
  \label{eq:disprho}
\left.-\vxi\bdot\grad\rho\right|_{R=\Rp-h}\approx \frac{N\rho(h)}{h}\xi_R(0)
\approx \frac{N\rho(h)}{h}\frac{\psi(0)}{g}\,,
\end{equation}
where we have evaluated $\xi_R(0)$ from the free boundary condition
(\ref{eq:freebc}) in the last step, taking $\Phi_1=0$ as appropriate
for a free mode of oscillation with negligible self-gravity.
Evidently, the neglected contribution (\ref{eq:disprho}) is comparable
to the total (\ref{eq:eulerianpert}), and therefore is not negligible
within the first node.

In Wu's powerlaw-sphere model, where
\begin{equation}
  \label{eq:wusphere}
\rho\propto (R_{\max}^2-R^2)^\beta
\end{equation}
the nodal depth is $\propto R/n$, except near the critical
colatitudes $\theta=\cos^{-1}(\pm\omega/2\Omega)$ (the ``singularity
belt'' in Wu's parlance) where it scales $\propto R/n^2$.  
Here $n=n_1+n_2$ in terms of Wu's modal indices
$n_i\ge0$, which are roughly proportional to the WKB wavenumbers of the
inertial modes: that is, $n\gg 1$ for shortwavelength modes.
Wu finds that the tidal forcing, which is proportional to the overlap
integral between the perturbing potential $\Phi_1$ and the modal
eigenfunction, scales with this index approximately as
$n^{-2\beta-1}$.  This is consistent with the idea that the excitation
occurs within the first node from the surface (and probably also
within the singularity belt) since the fraction of the planetary mass
within depth $2h$ of the surface for the density profile
\eqref{eq:wusphere} is $\Delta M/M \propto(2h)^{\beta+1}$.

It is true that the horizontal components of the velocity
and displacement near the boundary are larger than their radial
components by a factor $\sim\omega_{\rm dyn}^2/\Omega^2\gg1$, where
$\omega_{\rm dyn}\equiv(3g/R)^{1/2}$ is the dynamical frequency of the
planet. So it is likely that the errors in modal energies and
eigenfrequencies caused by the approximate boundary condition $\xi_R=0$
are slight.  But the overlap integrals are of a higher order of
smallness in $(\Omega/\omega_{\rm dyn})^2$, so that their relative errors
could be large.

The discussion so far does not make clear the sign of the
error (if there is one): perhaps the tidal coupling would be
\emph{larger} with the exact boundary condition.  The following
model system, which is borrowed from \cite{GNG87}, suggests that the
error is in the direction of overestimating the coupling.

Let the equation of state again be polytropic, with $\rho\propto w^N$,
$P\propto w^{N+1}$, and $\cs^2=w/N$, where $w\equiv(N+1)P/\rho$ is the
enthalpy, which remains finite in the incompressible limit $N\to 0^+$.
To match (\ref{eq:wusphere}), the unperturbed enthalpy should be
\begin{equation}
  \label{eq:w0}
  w(x,y,z)= w(0)\left(1-\frac{x^2+y^2+z^2}{R_{\max}^2}\right),
\end{equation}
and $N=\beta$.  Hydrostatic equilibrium in the
corotating frame (where $\vv=0$) requires
\begin{equation}
  \label{eq:hydrostatic}
0=  \grad\left[w+\Phi-\frac{1}{2}\Omega^2(x^2+y^2)\right],
\end{equation}
where $\Phi$ is the unperturbed potential.  Since $w$ and the
centrifugal term are quadratic functions, $\Phi$ must also be such a
function. In this regard, the model differs from the one considered by
Wu.  In effect, she calculates the potential due to the density
profile (\ref{eq:wusphere}) from Poisson's equation and uses this to
determine the enthalpy.  Consequently her pressure profile is not
simply a power of the density profile, and $\Gamma_1\equiv (\pd\ln
P/\pd\ln\rho)_S$ is not constant in her models, at least not for
$\beta>0$.  However, she argues at several points\footnote{For
  example, in \citet[after eq.~(C7)]{Wu05b}: ``The results only depend
  on the boundary behavior of $f(\Theta)$ as long as it is
  sufficiently smooth.  This explains why models with different
  polytrope representations ($\rho\propto[1-(r/R)]^\beta$ or
  $p\propto\rho^{1+1/\beta}$) give rise to essentially the same
  overlap integrals.''} that the tidal forcings she calculates depend,
at least for a smooth density profile, only on the behavior near the
boundary, where it is indeed approximately true that
$P\propto\rho^{(\beta+1)/\beta}$.  As a matter of fact, because of her
neglect of the $\cs^{-2}$ term in eq.~(\ref{eq:wavey}), the equation
of state doesn't enter her calculations of the normal modes: only the
density profile does, which could result from many isentropic
equations of state paired with an appropriate background potential.
The overlap integrals do depend upon the equation of state via the
sound speed, but as long as $\cs^2$ approaches zero linearly near the
boundary, it is hard to see how the overlap integrals for
short-wavelength modes could be sensitive to the full functional forms
of $\cs^2(R)$, and therefore of $\Phi_0(R)$, if they are excited near
the boundary.  For these reasons, it does not seem crucial that the
unperturbed potential be fully consistent with the mass distribution.

After elimination of the density in favor of the enthalpy, the
linearized equations become
\begin{subequations}\label{eq:polytide}
\begin{align}
\label{eq:pteuler}
-i\omega\vv+\boldsymbol{2\Omega\times v}+\grad w_1  &= -\grad\Phi_1,\\
\label{eq:ptcont}
-i\omega w_1 +\vv\bdot\grad w+N^{-1}w\diver\vv &=0.
\end{align}
\end{subequations}
Now suppose that the tidal potential is quadrupolar: that is, a
homogeneous and harmonic quadratic polynomial in $(x,y,z)$; for
definiteness,
\begin{equation}
  \label{eq:htide}
  \Phi_1 = \frac{A}{2}(x+iy)^2 e^{-i\omega t}
\end{equation}
where $A$ is a constant, and the exponential factor will be taken as
read hereafter.  Then since $w$ is also a second-degree polynomial,
eqs.~(\ref{eq:polytide}) can be satisfied by taking the components of
$\vv$ to be polynomials of the first degree, and $w_1$ of the second
degree.  After some algebra,
\begin{align}\label{eq:ptsol}
  v_x &= -i v_y = a(x+iy), & v_z&=0, & w_1= \frac{B}{2}(x+iy)^2,\nonumber\\
\mbox{where}~~a&= \frac{-i\omega A}{\omega(\omega+2\Omega)-4w(0)/R_{\max}^2}\,,
&\mbox{and}~B &= \frac{4iw(0)}{\omega R_{\max}^2}a\,.
\end{align}
Since $\diver\vv=0$, the equation of state (i.e. $N$) doesn't enter
the solution (\ref{eq:ptsol}).  Also, the relative vorticity
$\boldsymbol{\nabla\times}\vv=0$, so this solution has the same total
vorticity $2\vOmega$ as it would have in the absence of the tide and
therefore might be the solution of an initial value problem in which
the tide was ``turned on'' slowly.  The denominator in the expression
for $a$ vanishes at $\omega=-\Omega\pm(\omega_0^2-\Omega^2)^{1/2}$,
where $\omega_0\equiv 2\sqrt{w(0)}/R_{\max}$ is the dynamical
frequency of the model and therefore presumably $\gg\Omega$.  At the
unperturbed surface where $w=0$, eq.~(\ref{eq:ptcont}) reduces to the
free boundary condition that the lagrangian enthalpy perturbation
vanishes.

These details aside, the important point is that {\it the tidal
  response is entirely long-wavelength for any polytropic index when a
  free rather than rigid outer boundary condition is used,} at least
in this idealized coreless model, which uses a nonselfconsistent but
smooth unperturbed potential, and at least for a nonsynchronous body
in a circular orbit.  Short-wavelength inertial modes are not tidally
forced even though the fluid is compressible.

\subsection{Rigidity of the core}\label{subsec:rigidity}

OL04 assumed the core to be solid, and therefore impenetrable by low-frequency
but short-wavelength disturbances such as inertial waves, but sufficiently
plastic as to comply with the large-scale equilibrium tide as if it were fluid.
Here we re-examine the strength and solidity of the core.  In agreement
with OL04, we find that the elastic strength of even a solid core would be
negligible as regards the equilibrium tide.  However, we evaluate the
equilibrium tide differently than OL04, and we allow for the
density contrast between the core and its immediate surroundings.
Furthermore, we estimate that the core is most probably fluid rather than solid.

It is presumed that the cores of Jovian planets consist of elements
heavier than hydrogen and helium, more specifically of some
combination ``rock'' (refractory minerals such as silicates and iron)
and ``ice'' (molecular species such as ${\rm H_2O}$, ${\rm CH_4}$, and
${\rm NH_3}$) \citep{Guillot05}.  To support shear stress, these materials
would have to be in a solid phase.  The pressure at the surface
of the core is comparable to the central pressure of an $N=1$ polytrope:
\begin{equation}
  \label{eq:Pcore}
  P_{\rm c}\approx \frac{\pi}{8}\frac{G\Mp^2}{\Rp^4}\approx 4\times 10^{13}
\left(\frac{\Mp}{M_{\rm J}}\right)^2\left(\frac{\Rp}{R_{\rm J}}\right)^{-4}\ 
{\rm dyn\,cm^{-2}}\,,
\end{equation}
or $\sim 40\,{\rm Mbar}$.  The temperature is more difficult to
predict.  Jovian planets are supported mainly by degeneracy pressure,
so the central temperature has only a modest effect on the planetary
radius.   The temperature depends upon the planet's age and rate of cooling,
which in turn depend upon on uncertain opacities in the envelope, not
to mention the possibility of internal heating by tides.
Present estimates are in the range 
\begin{equation}
  \label{eq:Tcore}
T_{\rm c}\sim 2\mbox{-}4\times 10^4\,{\rm K}
\end{equation}
based on standard models fit to radii of transiting planets and the
estimated ages of their host stars \citep[e.g.][]{Arras_Bildsten06}.

The pressure (\ref{eq:Pcore}) is large compared to bulk moduli of
common refractory materials at room temperature---e.g. $K=1.7$ and
$1.0$~Mbar for iron and silicon, respectively---so rocky cores will be
compressed to densities $\gtrsim 20\,{\rm g\,cm^{-3}}$ [we derive this
number from a generic equation of state for ``rock'' by
\cite{Hubbard_Marley89}], i.e. a factor 3-10 times larger than their
densities at atmospheric pressure.  Under standard conditions, the
shear moduli of such materials are comparable to
their bulk moduli ($\mu_{\rm Fe}\approx \mu_{\rm Si}\approx 0.8\,{\rm Mbar}$).
Under compression, the bulk moduli rise more quickly than the shear
moduli because the former is associated with the degeneracy pressure
of the electrons, whereas the latter is a Coulomb effect having to do
with the ion lattice; for very large compression factors, one
therefore expects $\mu/K\propto (\rho/\rho_0)^{1/3}$.  With these
scalings, we can compare the elastic energy of the core under its
distortion by an equilibrium tide with the corresponding gravitational
energy.  Approximating the core by a constant density $\rho_{\rm c}$,
we have
\begin{eqnarray}
  \label{eq:distortions}
  \Delta E_{\rm grav}&=&\frac{k_p}{(1+2k_p)^2}\frac{G\Mc^2}{\Rc}\sigma^2,
\quad \Delta E_{\rm elas}= \frac{3}{4}\frac{\mu\Mc}{\rho_{\rm c}}\sigma^2,\nonumber\\
\frac{\Delta E_{\rm elas}}{\Delta E_{\rm grav}} &\approx& \frac{25\mu\Rc}{4G\Mc\rho_{\rm c}}
\,,
\end{eqnarray}
where $\sigma$ is the radial strain $\xi_{r,\max}/\Rc$ caused by the
quadrupolar tide.  The elastic modulus in the core's compressed state
should be $\mu\approx(\rho/\rho_0)^{1/3} \mu_0\approx 2\mu_0$, where the
subscript ``0'' refers to atmospheric conditions, hence no more than a
few Mbar, whereas $G\Mc\rho_{\rm c}/\Rc\approx 200\,{\rm Mbar}$ if
$\Mc=30\,{\rm M}_{\oplus}$ and $\rho_{\rm c}=20\,{\rm g\,cm^{-3}}$,
which is of course comparable to the pressure (\ref{eq:Pcore}).  So
the elastic energy of the equilibrium tide in the core is only $5\%$
of the gravitational energy, and therefore the
large-scale tidal distortion of the core should be nearly the same as
for a fluid.  However, elastic strength could still be enough to
prevent the propagation of inertial waves inside the core, because the
tidal period is long compared to the crossing time of an elastic wave
in the core, which is of order half an hour for the numbers above.

The discussion so far has been based on a solid core, but a liquid one
may be more likely.  Melting of an ionic lattice tends to occur at
$\Gamma\equiv (Z_{\rm eff}e)^2/r_i kT\sim 10^2$, a dimensionless
measure of the relative importance of Coulomb to thermal energies;
here $r_i\approx(3\rho/4\pi m_i)^{1/3}$ is the mean distance between
ions in terms of their mass, $m_i$ \citep[e.g.][and references
therein]{Shapiro_Teukolsky83}.  The question is what to use for the
effective charge $Z_{\rm eff}$ governing the ionic interactions.  For
silicon at its reference density of $2.33\,{\rm g\,cm^{-3}}$, for
example, this formula would predict a melting temperature $\approx
2\times 10^5(10^{-2}\Gamma)^{-1}\,{\rm K}$ if one were to take $Z_{\rm
  eff}=14$, the full charge on the nucleus.  In fact, most of that
charge is shielded by electrons whose orbits are much smaller than
$r_i$, so that $Z_{\rm eff}\approx 1$ is a more reasonable choice;
this leads to $990\Gamma_2^{-1}\,{\rm K}$, which is comparable to the
actual value, $T_{\rm m}=1693\,{\rm K}$.  As far as we know, there is
no experimental measurement of the melt temperature near $44\,{\rm
  Mbar}$, but if one simply scales it from 1~bar by the the reciprocal
of the inter-ionic distance, assuming that the Coulomb interaction is
characterized by a constant $Z_{\rm eff}$ throughout this range, then
$T_{\rm m}\,@\,20\,{\rm g\,cm^{-3}}\approx 3500\,{\rm K}$.  Despite
the crudeness of this argument, it therefore seems likely that
``rock'' should be molten at the much higher temperatures in
eq.~(\ref{eq:Tcore}).

To recap, the core is probably fluid, and even if it is solid, its
elastic strength will likely be unimportant for the equilibrium tide.
However, the density contrast between the high-$Z$ core and its surroundings
will affect its tidal distortion.  For simplicity, consider the
equilibrium tide in a nonrotating ``planet'' composed of two incompressible
fluids having different densities: $\rho=\rho_2$ in the core, $0\le R<\Rc$,
and $\rho=\rho_1<\rho_2$ outside it, $\Rc<R<\Rp$.  As usual, the perturbing
tidal potential is quadrupolar, $\Phi_{1,\rm ext}= \epsilon R^2P_2(\cos\theta)$.
Since the vorticity vanishes
except at the interface between the two fluids, the displacements can
be assumed to be proportional to the gradient of a scalar,
$\vxi=\grad\chi$, where $\chi$ satisfies Laplace's equation and is
$\propto P_2(\cos\theta)$; $\chi$ is discontinuous
at the interface but its radial derivative must be continuous.
This idealized problem can then be worked out analytically by matching
the radial parts of $\chi$ and of $\Phi_{1,\rm self}$ across the interface.
Included among these conditions is that the perturbed interface should
remain an equipotential,
\begin{equation*}
  \frac{\partial\chi}{\partial R}= -\frac{\Phi_1}{G\Mc/\Rc^2}\quad\mbox{at}~R=\Rc,
\end{equation*}
which is analogous to the free boundary condition at the surface.  If
$\delta\equiv\rho_2/\rho_1\ge 1$ is the ratio of densities
and $\eta\equiv\Rc/\Rp\le 1$,
it can then be shown that the radial strain in the core
is related to the radial strain at the interface by
\begin{equation}
  \label{eq:strainrat}
  \left.\frac{\xi_R}{R}\right|_{R<\Rc} =
\frac{5+(\delta-1)\eta^3}{5\delta-3(\delta-1)(1-\eta^5)}
\,\left.\frac{\xi_R}{R}\right|_{R=\Rp}\,.
\end{equation}
In the limit $\eta\ll 1$ (recall that we expect $\Rc\lesssim 0.2\Rp$),
the factor on the righthand side reduces approximately to
$5/(2\delta+3)$.  Since $\delta\approx2\mbox{-}4$, the core distorts
substantially less than the main body of the planet, though it does
not remain perfectly spherical.  We would therefore still expect the
tide to generate short-wavelength inertial waves in a rotating planet
with a core, but compared to the estimate (\ref{eq:Qval}) made for a
rigid core, the tidal $Q$ would increase by a factor
$\approx[1-5/(2\delta+3)]^{-2}=[(2\delta+3)/2(\delta-1)]^2$, i.e. by one half
to one order of magnitude.

\section{Summary}\label{Summary}

Motivated by possible applications to short-period extrasolar planets,
and by past work by \cite{Ogilvie_Lin04} and by \cite{Wu05b}, we have
studied the dynamical tide in isentropic fluid bodies with and without
cores.  Our goal has not been to obtain precise numerical results for
realistic planetary structures, since these are still quite uncertain,
but rather to provide a simple yet semi-quantitative physical
explanation of why a rigid core should give rise to short-wavelength
inertial waves.  We do this essentially by a combination of WKB and
perturbation theory, in which the small parameters are (i) wavelength
over radius, and (ii) core radius over planetary radius.  The
essential element in our model is the non-specular reflection of
inertial waves at a surface.  Such reflections can cause dramatic
changes in wavelength when the surface is nearly perpendicular to one
of the directions of the wavevector allowed by the WKB dispersion
relation at the tidal frequency.  In other words, we emphasize
critical latitudes rather than wave attractors.  The spirit of our
analysis is very much more local and informal than that of most
previous work on modes and tides in rotating bodies; we hope that the
local approach will be accepted as complementary rather than
contradictory to global analyses.

In order to obtain an upper bound on the tidal $Q$ from our local
approach, we consider the production of waves so close to the critical
latitudes, and hence so short in wavelength, that they damp
nonlinearly after a single encounter with the core.  Our assumption is
that wave attractors, which would involve waves that encounter the
core and the surface repeatedly before damping, can only lower $Q$
further, as would the escape of inertial waves into the stably
stratified radiative zone near the surface.  We have also examined the
physical basis for the assumption of a rigid core.  We find that
rock would probably be molten at the central temperatures and
pressures expected for hot Jupiters, and that even if the core were in
a solid phase, it would be sufficiently plastic that its large-scale
tidal distortion would closely approximate that of a fluid.  However,
because of the core's self-gravity and higher density, it will distort
less than its surroundings, so that short-wavelength inertial waves
will still be excited.  Our upper bound to the tidal $Q$ is in the
range $10^7\mbox{-}10^8(\Rc/0.2\Rp)^{-5}$.

We have criticized the tidal calculations by \cite{Wu05b} and by
\cite{Ivanov_Papaloizou07} for coreless isentropic bodies.  We believe
that their results, though probably correct for bodies that are
confined within a rigid \emph{outer} boundary, overestimate the
excitation of short inertial waves---and therefore underestimate
$Q$---for the astrophysically relevant case of a free outer boundary.
We support our case in part by reference to explicit analytical
solutions of the tidal response in idealized coreless models, albeit
ones that are themselves not entirely realistic.  Since short inertial
waves are certainly prevalent among the linear modes of such bodies,
as shown vividly by \cite{Rieutord_Valdettaro97} and
\cite{Rieutord_etal01}, and since they \emph{can} be excited in some
circumstances (e.g. when a core is present), a general theorem
concerning the conditions under which short-wavelength modes can and
cannot be forced by external potentials would be desirable.

\vspace{20pt} We thank Gordon Ogilvie and Yanquin Wu for generously
commenting on a draft of this paper, though we do not suggest that
they endorse all of its conclusions.  We thank Adam Burrows for advice
on cores masses and high-pressure equations of state.  This work was
supported in part by the National Science foundation under grant
AST-0707373 (to JG), and an NDSEG Graduate Fellowship (to CL).

\begin{appendix}
\section{Appendix: Tidal Q for an eccentric, synchronous orbit}
Through first order in orbital eccentricity $e$,
the quadrupolar part of the tidal potential acting on a rotationally aligned but
not necessarily synchronous planet is
\citep[e.g.][]{Zahn77}
\begin{multline}
  \label{eq:ecntidefull}
  \Phi_{1,\rm ext}=
  \frac{GM_*}{a^3}R^2\sqrt{\frac{4\pi}{5}}\mbox{Real}\left\{
\frac{1}{2}Y_{2,0}(\theta)-\sqrt{\frac{3}{2}}\,Y_{2,-2}(\theta,\phi)
e^{-i2(\Omega-n)t}\right.\\
\left.+e\left[\frac{3}{2}Y_{2,0}(\theta)e^{-int}
+\frac{\sqrt{6}}{4}Y_{2,-2}(\theta,\phi)e^{-i(2\Omega-n)t}
-\frac{7\sqrt{6}}{4}Y_{2,2}(\theta,\phi)e^{-i(3n-2\Omega)t}\right]\right\}.
\end{multline}
Here the azimuthal coordinate $\phi$ corotates with the planet.
Spherical harmonics rather than Legendre functions have been used,
to clarify the relative strengths of the various components.  For
circular orbits and non-synchronous spins, only the second term in
curly braces contributes to dissipation, because
the rest vanish or are constant in time.  This appendix is devoted to
synchronous ($n=\Omega$) but slightly noncircular cases,
$0<e\ll 1$.  Then all of the variable parts of the perturbing potential
are $\propto e$ and have the same frequency:
\begin{multline}
  \label{eq:ecntide}
  \Phi_1=
  e\frac{fGM_*}{a^3}R^2\sqrt{\frac{4\pi}{5}}\mbox{Real}\left\{\left[
\frac{3}{2}Y_{2,0}(\theta)
+\frac{\sqrt{6}}{4}Y_{2,-2}(\theta,\phi)
-\frac{7\sqrt{6}}{4}Y_{2,2}(\theta,\phi)\right]e^{-int}\right\}\,,
\end{multline}
where the factor $f$ accounts for the self-gravity of the equilibrium
tide as in eq.~\eqref{eq:synctide}. Following the prescription of
\S\ref{subsubsec:corescatt}, the corresponding amplitudes
of the incoming component of the equilibrum response at the core are
\begin{equation}
A_{\rm in}^{(m)}=-\frac{9f}{16}\frac{\Omega^2}{8\pi G\rho}
\frac{GM_*\Rc^2}{a^3} ~\times~
\begin{cases}
\frac{1}{2}e & m=-2\\
\frac{21}{2}e & m=+2\\
\vphantom{\frac{21}{2}}e & m=0.
\end{cases}
\end{equation}
Notice that the $m=0$ component is the same, apart from the factor of
eccentricity, as that of the $|m|=2$ component in the nonsynchronous circular case.
The radial energy flux of outgoing waves near the critical latitudes at the core is
\begin{equation}
\label{eq:FoutEc}
  \mathcal{F}_R\approx \frac{\rho}{\Rc}\sum\limits_m
\frac{\left|A_{\rm in}^{(m)}\right|^2}
{\sqrt{4\Omega^2-\omega_m^2}}\approx
\frac{223}{2\sqrt{3}}e^2\left(\frac{3f M_*\Rp^3}{32\Mp a^3}\right)^2
\rho\Rc^3\Omega^3\,.
\end{equation}
The wave power (\ref{eq:power}) then becomes
\begin{equation}
\label{eq:powerEc}
\dot E(\Delta\theta)\approx 
\frac{2007\pi}{512} e^2 \left[(1+2k_{\rm p})\frac{\sigma(\Rc)}
  {\sigma(\Rp)}\right]^2\left(\frac{M_*}{\Mp}\right)^2
\left(\frac{\Rp}{a}\right)^6\rho(\Rc)\Rc^5\Omega^3\Delta\theta.
\end{equation} 
The sum of the maximum potential energies in the tidal
distortions associated with each of the variable components in eq.~\eqref{eq:ecntide}
is
\begin{equation}\label{eq:deltaEecc}
\Delta E\approx 21e^2\,\frac{k_{\rm p} GM_*^2\Rp^5}{a^6}\,.
\end{equation}
Therefore, using the same approximations that led to eq.~\eqref{eq:Qval}, the tidal
$Q$ is
\begin{equation}\label{eq:Qecc}
Q \approx 2.1\times 10^7
\left(\frac{P}{4\,{\rm d}}\right)^2 
\left(\frac{\Mp}{\MJ}\right)\left(\frac{\Rp}{\RJ}\right)^2
\left(\frac{\Rc}{0.2\RJ}\right)^{-5}
\left(\frac{\Delta\theta}{0.1\,{\rm rad}}\right)^{-1}\, .
\end{equation}
This is very close to the result \eqref{eq:Qval} [but slightly
different by virtue of the rational numbers entering
eqs.~\eqref{eq:power}, \eqref{eq:deltaEsyn}, \eqref{eq:powerEc}, and
\eqref{eq:deltaEecc}] because both the synchronization and
circularization tides are dominated by one of their harmonic
components.
\end{appendix}


\clearpage

\end{document}